\documentclass[12pt,a4paper]{article}
\usepackage{axodraw}

\makeatletter
\newcommand\opm[2][]{{\bfseries\ifx#1\@mpty\else\textsf{#1:}\space\fi #2}}
\newcommand\eqref[1]{(\ref{#1})}

\DeclareSymbolFont{AMSb}{U}{msb}{m}{n}
\DeclareSymbolFontAlphabet{\mathbb}{AMSb}
\newcommand\inv{^{\raise.15ex\hbox{${\scriptscriptstyle -}$}\kern-.05em 1}} 
\newcommand{\al}{\alpha}

\newcommand\ket[1]{\vert #1 \rangle}
\newcommand\vev[1]{\langle #1 \rangle}                  
\newcommand\Vev[1]{\Bigl\langle #1 \Bigr\rangle}        
\newcommand\ext[1][]{\mathop{\raisebox{.2ex}{$\textstyle\bigwedge$}^{#1}}}
\newcommand\del{\partial}                               
\newcommand\dA{\partial_{\!A}}                          
\newcommand\bdA{\bar\partial_{\!A}}                     
\newcommand\bDA{\bar\Delta_A}                           
\newcommand\e{\mathrm{e}}                               
\newcommand\dint[2][]{\mathop{\mathalpha{\int#1}#2}}    
\renewcommand\vec[1]{\underline{#1}}                    
\newcommand\set[1]{\mathbb{#1}}                         
\newcommand\C{\set{C}}                                  
\newcommand\R{\set{R}}
\newcommand\Z{\set{Z}}
\newcommand\opname[1]{\mathop{\kern\z@\mathrm{#1}}\nolimits}    
\newcommand\Tr{\opname{Tr}}                             
\newcommand\Hom{\opname{Hom}}                           
\newcommand\End{\opname{End}}                           
\newcommand\mod{\opname{mod}}                           
\newcommand\group[1]{\mathop{\kern\z@\mathrm{#1}}\nolimits}     
\newcommand\SL{\group{SL}}
\newcommand\func[1]{\mathop{\kern\z@\mathsf{#1}}\nolimits}      
\newcommand\Hoch{\func{Hoch}}                           
\newcommand\Def{\func{Def}}
\newcommand\CO{\mathcal{O}}

\newcommand\CT{\mathcal{T}}
\renewcommand\section{\@startsection{section}{1}{\z@}%
                                    {-7ex \@plus -1ex \@minus -.2ex}%
                                    {2.5ex \@plus.2ex}%
                                    {\normalfont\large\scshape\centering}}                                   
\renewcommand\subsection{\@startsection{subsection}{2}{\z@}%
                                       {-5ex \@plus -1ex \@minus -.2ex}%
                                       {1.5ex \@plus.2ex}%
                                       {\normalfont\normalsize\scshape}}
\newcommand\ack{\section*{\ackname}}                            
\newcommand\ackname{Acknowledgements}
\newcommand\sectionname{}

\renewcommand\@seccntformat[1]{\ignorespaces\csname #1name\endcsname\space
                               \csname the#1\endcsname.\quad}   
\renewcommand\appendix{\par
  \setcounter{section}{0}%
  \setcounter{subsection}{0}%
  \renewcommand\thesection{\@Alph\c@section}
  \renewcommand\sectionname{\appendixname}}
\newdimen\captionmargin 
\newdimen\captionindent 
\newdimen\captionwidth 
\newcommand\captionfont{\slshape}
\newcommand\@captionlabel[1]{\textsc{#1:}\space}
\long\def\@makecaption#1#2{%
  \vskip\abovecaptionskip  
  \captionwidth\hsize 
  \advance\captionwidth -2\captionmargin
  \sbox\@tempboxa{\@captionlabel{#1}\captionfont #2}%
  \ifdim \wd\@tempboxa >\captionwidth
    \ifdim\captionindent>\z@ 
      \advance\captionwidth -\captionindent
      \hskip\captionindent
    \fi
    \hskip\captionmargin
    \parbox[t]{\captionwidth}{\leavevmode\hskip-\captionindent
      \@captionlabel{#1}\captionfont #2}%
  \else
    \global \@minipagefalse
    \hb@xt@\hsize{\hfil\box\@tempboxa\hfil}%
  \fi
  \vskip\belowcaptionskip}
\def\eqnarray{%
   \stepcounter{equation}%
   \def\@currentlabel{\p@equation\theequation}%
   \global\@eqnswtrue
   \m@th
   \global\@eqcnt\z@
   \tabskip\@centering
   \let\\\@eqncr
   $$\everycr{}\halign to\displaywidth\bgroup
       \hskip\@centering$\displaystyle\tabskip\z@skip{##}$\@eqnsel
      &\global\@eqcnt\@ne$\;\hfil{##}$\hfil
      &\global\@eqcnt\tw@$\;\displaystyle{##}$\hfil\tabskip\@centering
      &\global\@eqcnt\thr@@ \hb@xt@\z@\bgroup\hss##\egroup
         \tabskip\z@skip
      \cr
}
\makeatother
\begin{document}

%
%
\thispagestyle{empty}

\begin{flushright}\scshape
SPIN-00/15, ITP-UU-00/18, UG-00-07\\
hep-th/0006120\\ 
June 2000
\end{flushright}
\vskip10mm

\begin{center}

{\LARGE\scshape
Deformations of Topological\\ Open Strings
\par}
\vskip15mm

\textsc{Christiaan Hofman$^{1,2}$ \textnormal{and} Whee Ky Ma$^{1,3}$}
\par\bigskip
{\itshape
${}^1$Spinoza Institute, University of Utrecht,\\ 
      Leuvenlaan 4, 3508 TD Utrecht,
\par\medskip
${}^2$Institute for Theoretical Physics, University of Utrecht,\\ 
      Princetonplein 5, 3584 CC Utrecht,
\par\medskip
${}^3$Institute for Theoretical Physics, University of Groningen,\\
      Nijenborgh 4, 9747 AG Groningen
}\par\bigskip
\texttt{C.M.Hofman@phys.uu.nl, W.K.Ma@phys.rug.nl} 
\end{center}

\section*{Abstract}

Deformations of topological open string theories are described, with
an emphasis on their algebraic structure. They are encoded in the
mixed bulk-boundary correlators. They constitute the Hochschild
complex of the open string algebra -- the complex of multilinear maps
on the boundary Hilbert space. This complex is known to have the
structure of a Gerstenhaber algebra (Deligne theorem), which is also
found in closed string theory. Generalising the case of function
algebras with a $B$-field, we identify the algebraic operations of the
bulk sector, in terms of the mixed correlators. This gives a physical
realisation of the Deligne theorem. We translate to the language of
certain operads (spaces of $d$-discs with gluing) and $d$-algebras,
and comment on generalisations, notably to the AdS/CFT correspondence.
The formalism is applied to the topological A- and B-models on the
disc.

\newpage
\setcounter{page}{1}
%
%

\section{Introduction}

Open string theory in the presence of a constant background $B$-field
has received much attention recently (see e.g.\ \cite{codo,cafe,scho,seiwit}). 
The effect of the 2-form $B$-field is to deform the algebra of functions to an
algebra with a noncommutative associative star product. Kontsevich
\cite{kon1} has shown that associative star products are -- up to a
suitable gauge equivalence -- in one-to-one correspondence with
bivector fields, hence with constant 2-form background fields (the problem 
of deformation quantisation). Cattaneo and Felder \cite{cafe} have
demonstrated that the mapping effecting this correspondence has an
interpretation as the perturbative expansion of the path integral of a
two-dimensional topological $\sigma$-model on a disc. 
Furthermore, Kontsevich \cite{kon2} has reframed and
generalised deformation quantisation in the language of operads. 
These are general moulds for algebraic structures. 

The primary purpose of this paper is to understand these algebraic
structures in the general context of topological open string theory.

In Section \ref{tos} a short review of topological closed strings is
given, followed by a discussion of open string and mixed correlators,
the latter intertwining between the boundary and the bulk. Using Ward
identities we identify the mixed correlators with open string
deformations, and using factorisation we formulate algebraic relations
on them.

In Section \ref{hodef} we identify the algebraic (Hochschild)
structure of the deformations as that of a Gerstenhaber
\cite{zwie, kvz, ksv, stasheff, liz, wizwi}.

The relations of Section \ref{tos} show compatibility of these
structures between the deformations of the open string theory and the
bulk theory. This gives a physical realisation of the Deligne theorem,
which states that the deformation theory of an associative algebra has
the structure of a `two-dimensional field theory'.

The Deligne theorem is discussed further in Section \ref{ope}. The
structure of the deformation theory is linked to the operad of little
discs, which naturally describes open and closed string diagrams.

In Section \ref{abmodel} the formalism is applied to the topological
A- and B-models. This is rather interesting and it offers plenty of
scope for further research.

In Section \ref{concl} we mention some directions for further work.
Generalisations to higher dimensions can be made, being possibly
relevant to deformations of $M$-theory due to a background $C$-field.
Applications to open/closed string duality and AdS/CFT correspondence
also seem to have prospects.

\section{Algebraic Structure of Topological Open Strings}
\label{tos}

In this section we describe topological open and closed string theories, 
emphasising the mixed correlation functions.

\subsection{Review of Closed Strings}

Topological string theory is characterised by the existence of a
BRST operator $Q$, which is nilpotent, $Q^2=0$. The stress-energy
tensor is BRST-exact, $T_{\mu\nu} =\{Q,G_{\mu\nu}\}$, which implies
that the correlation functions of $Q$-closed operators are independent
of the metric on the worldsheet. There is a conserved quantum number 
called ghost number, such that the BRST operator has ghost number 1.  
The Hilbert space of topological closed string theory is spanned by 
operators $\phi_i$. We discuss the correlation functions 
of these operators on the sphere. The group of global conformal 
transformations on the sphere is $\SL(2,\C)$. Therefore, the ghost number 
anomaly is 6. This implies that the correlators on the sphere 
are nonvanishing only if the total ghost number of the operators is 6. 

The two-point functions define a metric on the space of operators,
\begin{equation}
\Vev{\phi_i\phi_j} = \eta_{ij},
\end{equation}
the three-point functions define `structure constants',
\begin{equation}
\Vev{\phi_i\phi_j\phi_k} = C_{ijk}
\end{equation}
The operators form an algebra, using the commutative OPE product 
\begin{equation}\label{clprod}
\phi_i\cdot\phi_j\sim C_{ij}{}^k\phi_k,
\end{equation}
where $C_{ij}{}^k=C_{ijl}\eta^{lk}$. 
Furthermore there are three-point functions
\begin{equation}
B_{ijk} = \Vev{ \oint\!\phi_i^{(1)}\phi_j\phi_k},
\end{equation}
where the contour integral goes around the insertion 
point of $\phi_j$.\footnote{In this paper we use the notation that a 
contour integral runs around the insertion point of the operator right behind it.
Note that for physical operators, having ghost number 2, these correlators vanish.} 
They define a bracket on the space of closed string operators
\begin{equation}\label{clhaak}
[\phi_i,\phi_j] = B_{ij}{}^k\phi_k,
\end{equation}
whose antisymmetry follows from the Ward identity.

One can construct a scalar `superfield', whose highest
component is denoted $\phi$ (a worldsheet scalar); the other components are
the descendants, denoted $\phi^{(1)}$ (a 1-form) and
$\phi^{(2)}$ (a 2-form). The components of the superfield satisfy the
descent equations,
\begin{eqnarray}
\{Q, \phi\} &=& 0,\\
\{Q, \phi^{(1)}\} &=& d\phi,\\
\{Q, \phi^{(2)}\} &=& d\phi^{(1)}. 
\end{eqnarray}
There are three types of observables, given by $\phi(P)$, $\oint_C\phi^{(1)}$, 
and $\int_\Sigma\phi^{(2)}$, where
$P$ is a point, $C$ is a contour and $\Sigma$ is the worldsheet surface. 
The first operators, inserted at a point, are local operators, while
the last type, as they are integrated over the worldsheet, can be used
to deform the action. The second type of operators are special for
two-dimensional local field theories. They can act on local operators,
by choosing the contour $C$ to enclose the insertion point $P$ of the
other operator.

\subsection{Correlators on the Disc}

On a worldsheet with boundary (disc) we have to consider the simultaneous
presence of closed string operators $\phi$ in the bulk and open string
operators $\alpha$ on the boundary. For the disc, the global conformal group 
is $\SL(2,\R)$, hence the ghost anomaly is 3. 
The correlators for the closed string operators are similar to the ones on 
the sphere discussed above. 
\begin{figure}
\begin{center}
  \begin{picture}(75,70)
    \CArc(40,25)(15,0,360)
    \Photon(10,25)(35,25)24
    \Vertex(35,25)1
    \Text(8,25)[r]{$\phi_i$}
    \Photon(45,25)(70,25)24
    \Vertex(45,25)1
    \Text(72,25)[l]{$\phi_j$}
    \Text(40,0)[m]{$\eta_{ij}$}
  \end{picture}
\qquad
  \begin{picture}(75,70)
    \CArc(40,25)(15,0,360)
    \Photon(10,5)(35,22)24
    \Vertex(35,22)1
    \Text(8,5)[r]{$\phi_i$}
    \Photon(45,22)(70,5)24
    \Vertex(45,22)1
    \Text(72,5)[l]{$\phi_j$}
    \Photon(40,55)(40,30)24
    \Vertex(40,30)1
    \Text(40,65)[m]{$\phi_k$}
    \Text(40,0)[m]{$C_{ijk}$}
  \end{picture}
\qquad
  \begin{picture}(75,70)
    \CArc(40,25)(15,0,360)
    \Photon(10,10)(35,20)24
    \LongArrowArc(40,20)(5,180,180)
    \Text(7,10)[r]{$\phi_i^{(1)}$}
    \Photon(40,20)(65,5)24
    \Vertex(40,20)1
    \Text(67,5)[l]{$\phi_j$}
    \Photon(40,55)(40,30)24
    \Vertex(40,30)1
    \Text(40,65)[m]{$\phi_k$}
    \Text(40,0)[m]{$B_{ijk}$}
  \end{picture}
\end{center}
\caption{Closed string interactions.}
\end{figure}
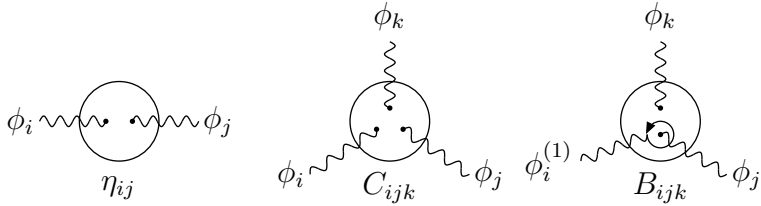
\begin{figure}
\begin{center}
  \begin{picture}(70,70)
    \CArc(35,25)(15,0,360)
    \Line(10,25)(20,25)
    \Vertex(20,25)1
    \Text(8,25)[r]{$\alpha_a$}
    \Line(50,25)(60,25)
    \Vertex(50,25)1
    \Text(62,25)[l]{$\alpha_b$}
    \Text(35,0)[m]{$G_{ab}$}
  \end{picture}
\qquad
  \begin{picture}(70,70)
    \CArc(35,25)(15,0,360)
    \Line(10,25)(20,25)
    \Vertex(20,25)1
    \Text(8,25)[r]{$\alpha_a$}
    \Line(50,25)(60,25)
    \Vertex(50,25)1
    \Text(62,25)[l]{$\alpha_b$}
    \Line(35,40)(35,50)
    \Vertex(35,40)1
    \Text(35,60)[m]{$\alpha_c$}
    \Text(35,0)[m]{$F_{abc}$}
  \end{picture}
\end{center}
\caption{Open string interactions.}
\end{figure}
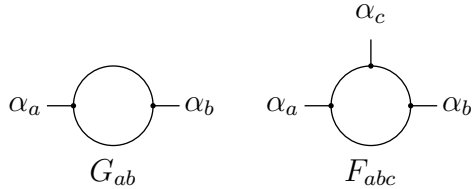
Then we consider correlators of boundary 
operators. The two-point functions, 
\begin{equation}
g_{ab} = \Vev{ \alpha_a\alpha_b },
\end{equation}
define a metric on the open string operators. 
The three-point functions, 
\begin{equation}
F_{abc} = \Vev{ \alpha_a\alpha_b\alpha_c },
\end{equation}
define  the structure constants of an algebra, 
the product of which we denote by $*$,
\begin{equation}
\alpha_a*\alpha_b = F_{ab}{}^c\alpha_c,
\end{equation} 
which is associative only on cohomology. Unlike in the closed string
algebra, the structure constants are not required to be symmetric, but
only cyclicly symmetric.

\begin{figure}
\begin{center}
  \begin{picture}(70,40)(0,-20)
    \CArc(35,25)(15,0,360)
    \Photon(10,25)(35,25)24
    \Text(8,25)[r]{$\phi_i$}
    \Vertex(35,25)1
    \Line(50,25)(60,25)
    \Vertex(50,25)1
    \Text(62,25)[l]{$\alpha_a$}
    \Text(35,0)[m]{$\Phi_{ia}$}
  \end{picture}
\qquad
  \begin{picture}(70,90)
    \CArc(40,45)(15,0,360)
    \Photon(10,55)(35,45)24
    \Vertex(35,45)1
    \Text(8,55)[r]{$\phi_i$}
    \Line(23,27)(30,34)
    \Vertex(30,34)1
    \Text(20,20)[r]{$\alpha_{a_1}^{(1)}$}
    \Line(57,27)(50,34)
    \Vertex(50,34)1
    \Text(55,22)[l]{$\alpha_{a_n}^{(1)}$}
    \Text(40,25)[m]{$\cdots$}
    \Line(40,60)(40,70)
    \Vertex(40,60)1
    \Text(40,80)[m]{$\alpha_{a_0}$}
    \Text(40,0)[m]{$\Phi_{ia_0a_1\ldots a_n}$}
  \end{picture}
\end{center}
\caption{Mixed interactions.}
\end{figure}
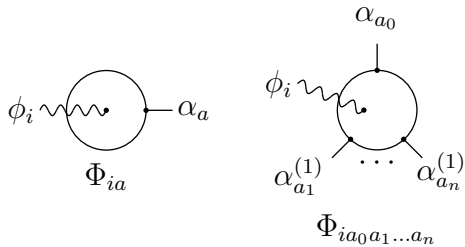
Finally, one has mixed correlators. First, there is a 
two-point function 
\begin{equation}
\Phi_{ia} = \Vev{\phi_i\alpha_a}.
\end{equation}
defining a map from the open to the closed string algebra, if we use
the closed string metric $\eta_{ij}$. Ghost counting guarantees
that the correlators discussed so far are all the correlators involving 
only scalars.

Next we insert integrated second descendants of bulk operators. 
Let $C_{ijk,l\ldots}$ denote the closed string correlator made from
the three-point function by inserting one additional second descendant
$\int\!\phi_l^{(2)}$. The Ward identity for the operator $G$ shows that 
these correlators are symmetric in all indices (WDVV equations)\cite{dvv,witeqn}. 
This implies that they are derivatives, e.g. $C_{ijk,l}=\del_l C_{ijk}$, 
with respect to certain formal parameters. Moreover, even the three-point functions 
are third derivatives of some function. Introducing coupling constants $t^i$ 
for each of the operators $\int\!\phi_i^{(2)}$, the symmetry allows us to define 
deformed closed string three-point functions by 
\begin{equation}
C_{ijk}(t) = \Vev{ \phi_i\phi_j\phi_k \e^{\,t^l\!\!\int\!\phi_l^{(2)}} }.
\end{equation}
In terms of these parameters, we can identify $\del_l=\del/\del t^l$.
In the deformed correlators, the extra insertions $t^l\int\!\phi_l$ can 
be viewed as a deformation of the bulk action.

Similarly, we introduce deformed open string three-point functions by 
\begin{equation}
F_{abc}(t) = \Vev{ \alpha_a\alpha_b\alpha_c \e^{\,t^l\!\!\int\!\phi_l^{(2)}} }.
\end{equation}
Derivatives with respect to $t$ again introduce extra bulk operators in the 
correlators. The parameters $t$ in this way deform the algebraic structure 
of the open string theory. 

Open string three-point functions deformed by open string insertions 
should naively be defined by 
\begin{equation}\label{opendef}
\tilde F_{abc}(s) = \Vev{ \alpha_a\alpha_b\alpha_c \e^{\,s^l\!\!\int\!\alpha_l^{(1)}} }.
\end{equation}
However, because of the ordering of operators on the boundary, 
this expression is not completely well-defined. Moreover, 
unlike in the closed string case, it is not 
possible to calculate open string correlators as derivatives 
of open-string three-point functions. For example, one has\footnote{Here and 
in the following integration of boundary operators will run between 
the insertion points of the neighbouring boundary operators. 
So the order of the boundary operators in the notation of the 
correlators always reflects the order of the operators on the boundary.}
\begin{equation}\label{noder}
\del_d \tilde F_{abc}(s=0) = \Vev{ \alpha_a\alpha_b\alpha_c
  \int_c^a\!\alpha_d^{(1)}}
 + \Vev{ \alpha_a\alpha_b\int_b^c\!\alpha_d^{(1)}\alpha_c }
 + \Vev{ \alpha_a\int_a^b\!\alpha_d^{(1)}\alpha_b\alpha_c }.
\end{equation}
The right hand side is well-defined, and it is the variation of
\eqref{opendef}, whatever the precise definition of the ordering. 

The separate higher-order open string correlators, defined by 
\begin{equation}
F_{a_0a_1\ldots a_n} = 
 \Vev{ \alpha_{a_0}\alpha_{a_1}\alpha_{a_2}\int\!\alpha_{a_3}^{(1)}\cdots\int\!\alpha_{a_n}^{(1)} },
\end{equation}
provide information which is in the $A_\infty$-structure of the open 
string theory (more about this below), and not in the associative structure. 
It is possible to choose an open string basis for which only the
(undeformed) three-point functions are nonzero; this corresponds to an
associative structure. However, when the open string theory is deformed 
this is not necessarily true, so that their deformed versions 
$F_{a_0a\ldots a_n}(t)$ do not vanish in general \cite{zwieoc}. 
The Ward identity for $G$ shows that these correlators are cyclic 
in the open string indices, rather than being symmetric. 
Similarly, higher-order mixed correlators 
\begin{equation}\label{twin}
\Phi_{ia_0a_1\ldots a_n} = 
\Vev{ \phi_i\alpha_{a_0}\int\!\alpha_{a_1}^{(1)}\cdots\int\!\alpha_{a_n}^{(1)} },
\end{equation}
are cyclic in the open string indices.

When we introduce extra closed string operators in the $\Phi$'s,
\begin{equation}
\Phi_{ija_0a_1\ldots a_n} =  
\Vev{ \phi_i\int\!\phi_j^{(2)}\alpha_{a_0}
      \int\!\alpha_{a_1}^{(1)}\cdots\int\!\alpha_{a_n}^{(1)} },
\end{equation}
we again find symmetry in $i$ and $j$ (see Appendix \ref{wrd}). 
As this is also true for the fully deformed correlators, satisfying 
$\Phi_{ija_0\ldots a_n} = \del_j\Phi_{ia_0\ldots a_n}$, they are integrable: 
there are functions $\Phi_{a_0\ldots a_n}$ such that 
$\Phi_{ia_0\ldots a_n} = \del_i\Phi_{a_0\ldots a_n}$. 
\begin{figure}
\begin{center}
\begin{picture}(190,90)
\put(0,0){
  \begin{picture}(70,90)
    \CArc(40,45)(15,0,360)
    \Photon(10,55)(35,45)24
    \Vertex(35,45)1
    \Text(7,57)[r]{$\phi_i^{(2)}$}
    \Line(23,27)(30,34)
    \Vertex(30,34)1
    \Text(25,22)[r]{$\alpha_{b}$}
    \Line(57,27)(50,34)
    \Vertex(50,34)1
    \Text(55,22)[l]{$\alpha_{c}$}
    \Line(40,60)(40,70)
    \Vertex(40,60)1
    \Text(40,80)[m]{$\alpha_{a}$}
    \Text(40,0)[m]{$\del_iF_{abc}$}
  \end{picture}
}
\put(90,40){=}
\put(120,0){
  \begin{picture}(70,90)
    \CArc(40,45)(15,0,360)
    \Photon(10,55)(35,45)24
    \Vertex(35,45)1
    \Text(8,55)[r]{$\phi_i$}
    \Line(23,27)(30,34)
    \Vertex(30,34)1
    \Text(22,20)[r]{$\alpha_{b}^{(1)}$}
    \Line(57,27)(50,34)
    \Vertex(50,34)1
    \Text(55,22)[l]{$\alpha_{c}^{(1)}$}
    \Line(40,60)(40,70)
    \Vertex(40,60)1
    \Text(40,80)[m]{$\alpha_{a}$}
    \Text(40,0)[m]{$\Phi_{iabc}$}
  \end{picture}
}
\end{picture}
\end{center}
\caption{Deformation of the product.}
\label{defpr}
\end{figure}
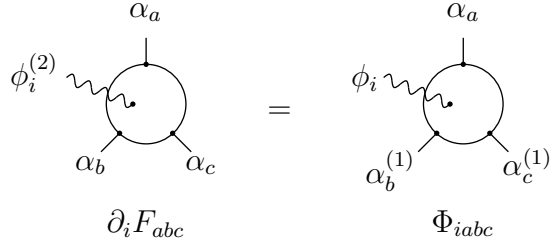
We can however go further, and show that the mixed correlators 
are related to deformations of the open string algebra. The derivation, 
which is in Appendix \ref{wrd}, is similar to the derivation of the 
WDVV equations \cite{dvv}. The result is
\begin{equation}\label{defprod}
\Vev{ \int\!\phi_i^{(2)}\alpha_a\alpha_b\alpha_c }
 = \Vev{ \phi_i\alpha_a\int\!\alpha_b^{(1)}\int\!\alpha_c^{(1)} },
 \quad\mathrm{i.e.}\quad \del_i F_{abc} = \Phi_{iabc}. 
\end{equation}
This expresses that the element $\Phi_{iabc}$ is a deformation of the
product by closed string operators, see Figure \ref{defpr}.

Upon inserting extra operators $\int\!\alpha_b^{(1)}$, we obtain 
$\del_i F_{a_0\ldots a_n} = \Phi_{ia_0\ldots a_n}$. This shows again that the
correlators $\Phi_{ia_0\ldots a_n}$ are derivatives, and it provides a
way to calculate the functions $\Phi_{a_0\ldots a_n}$.

\begin{figure}
\begin{center}
\begin{picture}(235,80)
\put(0,0){%
  \begin{picture}(70,80)(0,10)
    \CArc(40,45)(15,0,360)
    \Photon(10,55)(35,45)24
    \Vertex(35,45)1
    \Text(8,55)[r]{$\{\phi_i, Q\}$}
    \Line(23,27)(30,34)
    \Vertex(30,34)1
    \Text(25,22)[r]{${a_1}$}
    \Line(57,27)(50,34)
    \Vertex(50,34)1
    \Text(55,22)[l]{${a_n}$}
    \Line(40,60)(40,70)
    \Vertex(40,60)1
    \Text(40,78)[m]{${a_0}$}
    \Text(40,25)[m]{$\cdots$}
  \end{picture}
}
\put(90,30){$\displaystyle =\sum_k(-1)^k$}
\put(165,0){%
  \begin{picture}(70,90)(0,10)
    \CArc(40,45)(15,0,360)
    \Photon(10,65)(35,50)24
    \Vertex(35,50)1
    \Text(8,68)[r]{$\phi_i$}
    \Line(25,45)(15,45)
    \Vertex(25,45)1
    \Text(13,45)[r]{${a_1}$}
    \Line(55,45)(65,45)
    \Vertex(55,45)1
    \Text(67,45)[l]{${a_n}$}
    \Line(40,60)(40,70)
    \Vertex(40,60)1
    \Text(40,78)[m]{${a_0}$}
    \Text(22,35)[m]{$\cdot$}
    \Text(28,29)[m]{$\cdot$}
    \Text(58,35)[m]{$\cdot$}
    \Text(52,29)[m]{$\cdot$}
    \Line(40,30)(35,20)
    \Line(40,30)(45,20)
    \Vertex(40,30)1
    \Text(35,12)[r]{${a_k}$}
    \Text(45,12)[l]{${a_{k+1}}$}
  \end{picture}
}
\end{picture}
\end{center}
\caption{Factorisation of the BRST operator.}
\label{brst}
\end{figure}
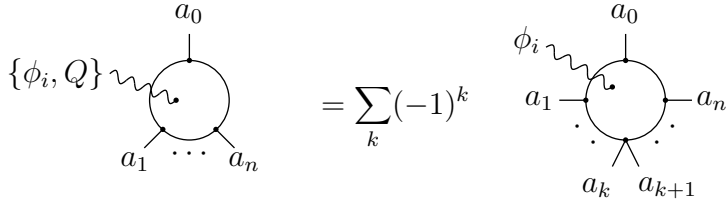

We want to view the mixed correlators as intertwiners between the closed 
string algebra and the deformations of the $A_\infty$ structure, given 
by the boundary correlators $F_{abc\ldots}$. An essential 
structure of the topological bulk theory is the BRST operator. 
A BRST operator acting on the closed string operator in the mixed 
correlators can be deformed to a contour around the boundary operators. 
Using the descent equations for the boundary operators gives the 
following identity, also depicted in Figure \ref{brst}.
\begin{eqnarray}\label{randtermen}
\Vev{ \{\phi_i,Q\}\alpha_{a_0}\int\!\alpha_{a_1}^{(1)}\cdots
  \int\!\alpha_{a_n}^{(1)} }
&=& \Vev{ \phi_i(\alpha_{a_0}*\alpha_{a_1})\int\!\alpha_{a_2}^{(1)}\cdots
  \int\!\alpha_{a_n}^{(1)} } \nonumber\\
&&+ (-1)^{n+1}\Vev{ \phi_i\int\!\alpha_{a_1}^{(1)}\cdots
  \int\!\alpha_{a_{n-1}}^{(1)}(\alpha_{a_n}*\alpha_{a_0})} \\
&&+ \sum_{k=1}^{n-1} (-1)^{k}
 \Vev{ \phi_i\alpha_{a_0}\int\!\alpha_{a_1}^{(1)}\cdots
  \int(\alpha_{a_k}*\alpha_{a_{k+1}})^{(1)}\cdots\int\!\alpha_{a_n}^{(1)} }. 
\nonumber
\end{eqnarray}
In this derivation, the boundary operators are taken on-shell (BRST-closed), 
while for $\phi_i$ we take an arbitrary local closed 
string operator. The boundary 
operators are assumed to have odd ghost degree; otherwise extra signs 
are introduced. 

We will  see that the factorisation of certain diagrams gives the
algebra of mixed correlators the full structure of a closed string
algebra. First, the closed string metric $\eta_{ij}$ factorises in the
open string channel,
\begin{equation}
\eta_{ij} = \Phi_{ia}g^{ab}\Phi_{jb}.
\end{equation}
There is also a straightforward 
factorisation of the $(n+2)$-point functions 
with two closed string scalars: 
\begin{equation}
\Vev{ \phi_i\phi_j\alpha_{b}\int\!\alpha_{a_1}^{(1)}\cdots\int\!\alpha_{a_n}^{(1)} }. 
\end{equation}
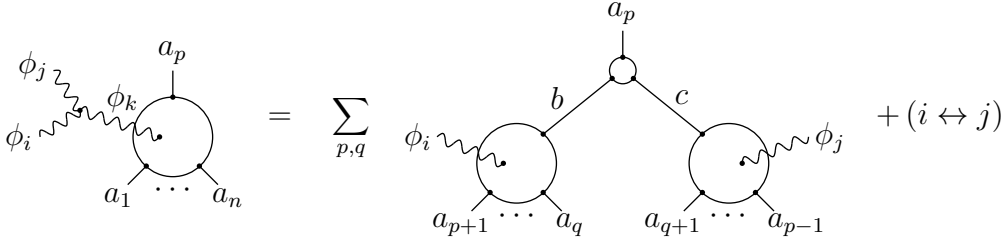
\begin{figure}
\begin{center}
\begin{picture}(380,120)
\put(0,10){
  \begin{picture}(85,90)(-10,0)
    \CArc(50,45)(15,0,360)
    \Photon(15,55)(45,45)25\Vertex(45,45)1
    \Vertex(15,55)1
    \Text(25,60)[l]{$\phi_k$}
    \Photon(15,55)(0,45)23
    \Text(-3,45)[r]{$\phi_i$}
    \Photon(15,55)(5,70)23
    \Text(3,72)[r]{$\phi_j$}
    \Line(33,27)(40,34)
    \Vertex(40,34)1
    \Text(35,22)[r]{${a_1}$}
    \Line(67,27)(60,34)
    \Vertex(60,34)1
    \Text(65,22)[l]{${a_n}$}
    \Line(50,60)(50,70)
    \Vertex(50,60)1
    \Text(50,77)[m]{${a_p}$}
    \Text(50,25)[m]{$\cdots$}
  \end{picture}}
\put(100,60){$=\quad\displaystyle\sum_{p,q}$}
\put(150,0){
  \begin{picture}(160,140)
    \CArc(80,80)(5,0,360)
    \Line(80,85)(80,95)
    \Vertex(80,85)1
    \Text(80,102)[m]{${a_p}$}
    \Line(76,77)(50,56)\Vertex(76,77)1\Vertex(50,56)1
    \Text(58,70)[r]{$b$}
    \Line(84,77)(110,56)\Vertex(84,77)1\Vertex(110,56)1
    \Text(100,70)[l]{$c$}
    \CArc(40,45)(15,0,360)
    \Photon(10,55)(35,45)24
    \Vertex(35,45)1
    \Text(8,55)[r]{$\phi_i$}
    \Line(23,27)(30,34)
    \Vertex(30,34)1
    \Text(30,22)[r]{${a_{p+1}}$}
    \Line(57,27)(50,34)
    \Vertex(50,34)1
    \Text(55,22)[l]{${a_q}$}
    \Text(40,25)[m]{$\cdots$}
    \CArc(120,45)(15,0,360)
    \Photon(150,55)(125,45)24
    \Vertex(125,45)1
    \Text(153,55)[l]{$\phi_j$}
    \Line(103,27)(110,34)
    \Vertex(110,34)1
    \Text(110,22)[r]{${a_{q+1}}$}
    \Line(137,27)(130,34)
    \Vertex(130,34)1
    \Text(135,22)[l]{${a_{p-1}}$}
    \Text(120,25)[m]{$\cdots$}
  \end{picture}}
\put(330,60){$+\,(i\leftrightarrow j)$}
\end{picture}
\end{center}
\caption{Factorisation of the product.}
\label{prfact}
\end{figure}
This gives the following identity:
\begin{equation}\label{produkt}
C_{ij}{}^k\Phi_{ka_0\ldots a_n} = \sum_{p,q=0}^{n} (-1)^{p+q-1}
 F_{bca_{p}} \Phi_i{}^{b}{}_{a_{p+1}\ldots a_q}\Phi_j{}^{c}{}_{a_{q+1}\ldots a_{p-1}} 
+ (i\leftrightarrow j).
\end{equation}
This is a commutative product on the algebra of physical operators, 
depicted in Figure \ref{prfact}. Notice that the integrated open string 
operators can in principle also be inserted at the open string three-point 
interaction, but as remarked before, this amounts to considering the 
$A_\infty$ rather than the associative structure. That, in turn, corresponds 
to the definition \eqref{ghpr} of the product of the Gerstenhaber algebra 
acquiring higher-order corrections to the bilinear $*$ operation.

\begin{figure}
\begin{center}
\begin{picture}(230,140)
\put(0,0){
  \begin{picture}(75,90)
    \CArc(50,45)(15,0,360)
    \Photon(25,65)(43,50)24
    \LongArrowArc(45,45)(5,120,120)
    \Text(25,70)[r]{$\phi_i^{(1)}$}
    \Photon(20,45)(45,45)24
    \Vertex(45,45)1
    \Text(17,45)[r]{$\phi_j$}
    \Line(33,27)(40,34)
    \Vertex(40,34)1
    \Text(28,22)[m]{$a_1$}
    \Line(67,27)(60,34)
    \Vertex(60,34)1
    \Text(65,22)[l]{$a_n$}
    \Line(50,60)(50,70)
    \Vertex(50,60)1
    \Text(50,77)[m]{$a_0$}
    \Text(50,25)[m]{$\cdots$}
  \end{picture}}
\put(90,50){$=\quad\displaystyle\sum_{p,q}\pm$}
\put(150,0){
  \begin{picture}(80,140)(0,20)
    \CArc(40,100)(15,0,360)
    \Photon(15,125)(35,105)24
    \Vertex(35,105)1
    \Text(13,127)[r]{$\phi_i$}
    \Line(50,111)(57,118)
    \Vertex(50,111)1
    \Text(40,132)[m]{$a_n$}
    \Line(15,100)(25,100)
    \Vertex(25,100)1
    \Text(57,123)[l]{$a_0$}
    \Line(40,115)(40,125)
    \Vertex(40,115)1
    \Text(13,100)[r]{$a_1$}
    \Line(23,82)(30,89)
    \Vertex(30,89)1
    \Text(25,77)[r]{$a_q$}
    \Text(22,94)[r]{$\cdot$}\Text(24,89)[r]{$\cdot$}
    \Line(57,82)(50,89)
    \Vertex(50,89)1
    \Text(55,77)[l]{$a_{p}$}
    \Text(60,105)[l]{$\vdots$}
    \Text(43,72)[l]{$b$}
    \CArc(40,45)(15,0,360)
    \Photon(10,45)(35,45)24
    \Vertex(35,45)1
    \Text(8,45)[r]{$\phi_j$}
    \Line(40,60)(40,85)\Vertex(40,60)1\Vertex(40,85)1
    \Line(23,27)(30,34)
    \Vertex(30,34)1
    \Text(30,22)[r]{$a_{q+1}$}
    \Line(57,27)(50,34)
    \Vertex(50,34)1
    \Text(55,22)[l]{$a_{p-1}$}
    \Text(40,25)[m]{$\cdots$}
  \end{picture}}
\end{picture}
\end{center}
\caption{Factorisation of the bracket.}
\label{brfact}
\end{figure}
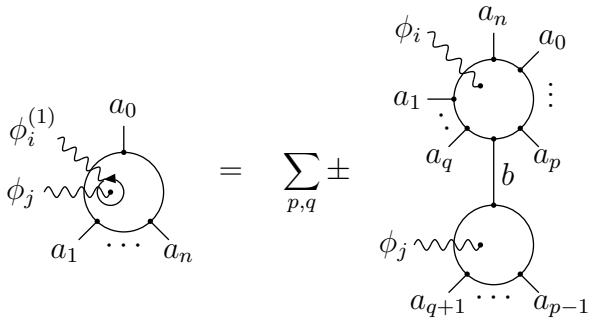

Finally, factorisation of the bracket,
\begin{equation}
\Vev{ \alpha_{a_0}\int\!\alpha_{a_1}^{(1)}\cdots 
  \int\!\alpha_{a_n}^{(1)}\oint\!\phi_i^{(1)}\phi_j },
\end{equation}
gives the identity
\begin{equation}\label{haakje}
B_{ij}{}^k\Phi_{ka_0a_1\ldots a_n} = 
\sum_{p,q} (-1)^{p+q-1} \Phi_{iba_{p}\ldots a_{q}}
 \Phi_{j}{}^b{}_{a_{q+1}\ldots a_{p-1}},
\end{equation}
depicted in Figure \ref{brfact}. We will prove this identity 
in the next section. 

\section{Deformations and Hochschild Complex}
\label{hodef}

The above structure of the BRST operator, product and bracket will be
interpreted as a natural algebraic structure on the deformations of
the open string theory.

\subsection{String Theories and Algebras up to Homotopy}

It has been shown that open string theory has the structure of a
homotopy associative, or $A_\infty$-algebra \cite{gerst,stas,wit,kon4}
(see Appendix \ref{hom}).  This is a generalisation of associative
algebra, where the associativity constraint is replaced by an
infinite number of similar relations.  We saw this reflected in the
fact that due to the ordering on the boundary, higher-point functions
cannot be written as derivatives, cf. eq. \eqref{noder}.

It has also been recognised that (topological) closed string theory is
related to homotopy Lie algebras, and with extra structure, to
(homotopy) Gerstenhaber algebras, see e.g.
\cite{zwie,kvz,ksv,stasheff, liz, wizwi}. A Gerstenhaber algebra is an
associative algebra with a symmetric product (eq. \eqref{clprod}),
which in addition is provided with a graded Lie bracket (eq.
\eqref{clhaak}) of (ghost) degree $-1$, satisfying certain associativity
and compatibility relations (see Appendix \ref{hom}).

More familiar in string theory is the notion of a BV algebra, which is
a Gerstenhaber algebra whose bracket can be expressed in terms of a BV
operator $\Delta$ of ghost number $-1$, squaring to 0. It is not a
derivation of the product though, as the failure of the Leibniz rule
determines the bracket,\footnote{For the bosonic string, the BV operator 
is given by the anti-ghost zero-mode $b_0^-=b_0-\bar b_0$.}
\begin{equation}
\{\phi_i,\phi_j\} = \Delta(\phi_i\cdot\phi_j) 
  - \Delta\phi_i\cdot\phi_j \mp \phi_i\cdot\Delta\phi_j. 
\end{equation}
A generalisation of the claims in the present paper to deformations of
closed strings would require a closer look at the differences between
Gerstenhaber and BV structures.

Gerstenhaber and BV algebras can be extended to homotopy Gerstenhaber
and homotopy BV algebras. Physically, the higher homotopies correspond
to BRST-exact terms; passing to BRST cohomology eliminates them.

\subsection{Gerstenhaber Structure}

How an algebra $A$ is deformed is determined by its Hochschild complex
$\func{Hoch}(A)$. Let us first focus on graded $A_\infty$-algebras
$A$, such as open string algebras. Besides the ordinary bilinear
operation (product), these can have $n$-linear operations, of degree
$2-n$, satisfying the master equation (see Appendix \ref{hom}).

For any algebraic structure, the relations and symmetries can be formulated 
in terms of multilinear maps. For example, the structure constants 
$F^{a_0}{}_{a_1\ldots a_n}$ of the $A_\infty$ algebra of the boundary 
operators define an $n$-linear map on the space of these operators. 
Also the deformations of the algebraic structure are multilinear maps. 
We found that the deformations of the structure constants are encoded in 
the mixed correlation functions \eqref{twin}. In fact, for a given closed
string operator $\phi_{i}$, we may identify the object in \eqref{twin} as
a map $\Phi:A^{\otimes n}\rightarrow A$, using the open string metric
$g_{ab}$:
\begin{equation}\label{corrhoch}
\Phi_i(\al_{a_1}, \ldots, \al_{a_n}) 
 = \Vev{ \phi_i\al_b \int\!\al_{a_1}^{(1)} \cdots \int\!\al_{a_n}^{(1)} } g^{bc} \al_c
 = \Phi_{iba_1 \ldots a_n} \al^b = \del_i F_{ba_1\ldots a_n} \al^b,
\end{equation}
with inverse relation
\begin{equation}\label{mixed}
\Phi_{iba_1 \ldots a_n} = \Vev{ \al_b\, \Phi_i(\al_{a_1},\ldots,\al_{a_n}) }.
\end{equation}
The last equality in \eqref{corrhoch} shows that we can interpret the maps
$\Phi_i$ as infinitesimal deformations of the higher structure
constants $F^a{}_{bc\ldots}$ of the $A_\infty$ algebra.

The vector spaces $C^n(A,A)=\Hom(A^{\otimes n},A)$, consisting of $n$-linear maps in $A$, 
define the degree $n$ space of the Hochschild complex $\Hoch(A)$ of the algebra $A$. 
Using the closed string metric $\eta_{ij}$ as well, one can equivalently 
view $\Phi$ as a map from $\Hom(A^{\otimes n},A)$ to the closed string algebra 
of the $\phi_i$. In this way, we have canonical maps between the Hochschild 
complex of the open string algebra and the space of closed string operators. 
We will usually identify the correlator with the corresponding mapping. 

The Hochschild complex of a (homotopy) associative algebra is naturally 
endowed with three operations: a coboundary operator (of degree 1), 
a product (of degree 0) and a bracket (of degree $-1$). 
\begin{itemize}
\item The Hochschild coboundary operator $\delta$ acts by contraction
and by adding two boundary terms. For $\Phi \in C^n(A,A)$ it 
is defined by
\begin{eqnarray}\label{hoch}
(\delta\Phi) (\al_1, \ldots, \al_{n+1}) &=& \al_1 * \Phi(\al_2, \ldots, \al_{n+1}) 
  + (-1)^{n+1} \Phi(\al_1, \ldots, \al_{n}) * \al_{n+1} \nonumber\\
&&+ \sum_{k=1}^{n} (-1)^{k} \Phi(\al_1, \ldots , \al_k * \al_{k+1}, \ldots, \al_{n+1}).
\end{eqnarray}
Equation \eqref{randtermen} expresses that this Hochschild coboundary operator 
equals the BRST operator $Q$ of the closed string,
\begin{equation}
\delta \Phi (\al_{a_1}, \ldots, \al_{a_{n+1}}) = 
 \Vev{ \{\Phi, Q\} \al_b \int\!\al_{a_1}^{(1)} \cdots \int\!\al_{a_{n+1}}^{(1)} } \al^b.
\end{equation}

\item The product is the cup product of two elements $\Phi_i \in C^{n_i}(A,A)$, defined by   
\begin{equation}\label{ghpr}
(\Phi_1 \cup \Phi_2)(\al_1, \ldots , \al_{n_1+n_2}) = 
(-1)^{n_1 n_2} \Phi_1(\al_1, \ldots, \al_{n_1}) * 
\Phi_2(\al_{n_1+1}, \ldots, \al_{n_1+n_2}).
\end{equation}
It is associative, but graded commutative only on cohomology. In 
deformed topological open string theory it is given by \eqref{produkt}. 

\item The (twisted) Lie bracket (the Gerstenhaber bracket) for 
$\Phi_i \in C^{n_i}(A,A)$ is given by
\begin{equation}\label{ghhaakje}
\{\Phi_1,\Phi_2\} = \Phi_1\circ\Phi_2 -(-1)^{(n_1-1)(n_2-1)} \Phi_2\circ\Phi_1,
\end{equation}
where $\circ$ is the composition of mappings on complexes,
\begin{eqnarray}
&&(\Phi_1 \circ \Phi_2)(\al_1, \ldots, \al_{n_1 + n_2-1}) = \\
 &&\qquad\sum_{k=1}^{n_1+n_2-1} (-1)^{k(n_2-1)} 
\Phi_1 (\al_1,\ldots,\al_k,\Phi_2(\al_{k+1},\ldots,\al_{k+n_2}),\al_{k+n_2+1},\ldots,\al_{n_1+n_2-1}).
\nonumber
\end{eqnarray}
We mentioned, and we will prove below that in deformed topological
open string theory it is given by the closed string bracket \eqref{haakje}.
\end{itemize}

The Hochschild coboundary operator and the Gerstenhaber bracket give
$C^*(A,A)$ the structure of a differential graded Lie algebra, with
the cup product it is also a differential Gerstenhaber algebra. The
master equation for the Gerstenhaber structure says that $Q$ is a
nilpotent derivation for the product and the bracket, and that the
bracket satisfies the Jacobi identity up to BRST-exact terms
(cf. \cite{stasheff}). These terms correspond to higher homotopies in the
Gerstenhaber algebra.

\subsection{Maurer-Cartan Equation and Bracket}

The Hochschild complex shifted by 1 of an associative algebra can be
regarded as its deformation complex. The associativity of the product
can be stated in the $A_\infty$ language as $d_2^2 = 0$.  The objects
which deform $A$ are in its Hochschild cohomology. The $\Phi$-deformed
product can be written 
$d_2'(\alpha_a, \alpha_b) = d_2(\alpha_a,\alpha_b) + \Phi(\alpha_a, \alpha_b)$, 
where $d_2(\alpha_a, \alpha_b) = \alpha_a * \alpha_b$ and
\begin{equation}\label{expansion}
\Phi(\alpha_a, \alpha_b) = (F_{ab}{}^c(t) - F_{ab}{}^c(0))\alpha_c 
 = t^i \Phi_{iab}{}^c\alpha_c + \frac{1}{2} t^it^j \Phi_{ijab}{}^c\alpha_c + \CO(t^3).
\end{equation}
Figure \ref{defpr} and eq. \eqref{defprod} show the
deformation caused by $\phi_i^{(2)}$ as a Hochschild cocycle.

Preserving associativity when deforming with a Hochschild
cocycle $\Phi$ is equivalent to the Maurer-Cartan equation for $\Phi$,
\begin{equation}\label{mc}
\delta\Phi + \Phi\circ\Phi = \delta\Phi + \frac{1}{2}\{\Phi,\Phi\}=0,
\end{equation}
because we can write \eqref{hoch} as $d_2\circ\Phi + \Phi\circ d_2 = \delta\Phi$.
For infinitesimal deformations $\Phi=t_i\Phi_i+\CO(t^2)$, we find $\delta\Phi=0$, 
therefore $\Phi$ is BRST-closed. In fact, the infinitesimal element is 
in the Hochschild cohomology. The Maurer-Cartan 
equation can be used to find a finite deformation in terms of the infinitesimal one. 
For example, for a single deformation parameter $t^i=t$, we may write 
\eqref{expansion} as $\Phi(t) = \sum_{n\geq1}\Phi_{(n)}t^n$, and 
we find recursive relations 
\begin{equation}
\delta\Phi_{(n)} = -\frac{1}{2}\sum_{k+l=n;~k,l\geq1}\{\Phi_{(k)},\Phi_{(l)}\}.
\end{equation}

We will  show that in deformed topological open string theory, the 
bracket $\{ \cdot, \cdot\}$ is related to the closed string bracket,
\begin{equation}
[\phi_i, \phi_j ] = \oint\!\phi_i^{(1)} \phi_j,
\end{equation}
contracted with the couplings. 
The Maurer-Cartan equation can be derived from the factorisation of the 
fully deformed four-point function on $\phi$-cohomology. Without integrated 
boundary insertions we write, keeping only the position of $\alpha_a$ free,
\begin{eqnarray}\label{defbracket}
\int d \Vev{ \al_a\al_b\al_c\al_d \,\e^{\,t^k\!\!\int\!\phi_k^{(2)}} }
 &=& \int \Vev{ \{Q, \al_{a}^{(1)}\} \al_b\al_c \al_d \,\e^{\,t^k\!\!\int\!\phi_k^{(2)}} } 
\nonumber\\
 &=& \Vev{ \int\!\al_{a}^{(1)} \al_b\al_c\al_d \Bigl\{Q, t^i\!\int\!\phi_i^{(2)} \Bigr\} \,\e^{\,t^k\!\!\int\!\phi_k^{(2)}} } 
\nonumber\\
 &=& \Vev{ \int\!\al_{a}^{(1)} \al_b\al_c\al_d  \,t^i\! \oint \phi_i^{(1)} \,\e^{\,t^k\!\!\int\!\phi_k^{(2)}} } 
\\
 &=& t^i t^j \Vev{ \int\!\al_{a}^{(1)} \int\!\al_{b}^{(1)}\int\!\al_{c}^{(1)} \al_d  [\phi_i,\phi_j] \,\e^{\,t^k\!\!\int\!\phi_k^{(2)}} }, 
\nonumber
\end{eqnarray}
where we used the descent equations and the fact that that the BRST operator
$Q$ and the operation $\oint\!\phi_i$ act as derivations on the exponential. 

The left-hand side can be written as a pure boundary sum, 
\begin{equation}\label{assoc}
\Vev{ \al_a\al_b\al^e \,\e^{\,t^k\!\!\int\!\phi_k^{(2)}} } 
 \Vev{ \al_e\al_c\al_d \,\e^{\,t^k\!\!\int\!\phi_k^{(2)}} } 
- \Vev{ \al_b\al_c\al^e \,\e^{\,t^k\!\!\int\!\phi_k^{(2)}} } 
   \Vev{ \al_e\al_d\al_a \,\e^{\,t^k\!\!\int\!\phi_k^{(2)}} }. 
\end{equation}
Associativity of the deformed product is the vanishing of this
expression. Putting \eqref{defbracket} to zero is exactly the
Maurer-Cartan equation \eqref{mc}, if we identify the bracket on the
open string deformations with the bracket in the closed string theory.

Up to first order in $t$, the vanishing is guaranteed by 
undeformed associativity and $\delta$-closedness of $\phi$.
In second order in $t$, we find from \eqref{defbracket} and \eqref{assoc}:
\begin{eqnarray}\label{brack}
2\Vev{ \int \al_a^{(1)}\int \al_b^{(1)}\int \al_c^{(1)}\al_d [\phi_i,\phi_j] } 
&=& \Vev{ \al_a\al_b\al^e \int\!\phi_i^{(2)}\int\!\phi_j^{(2)} } 
 \Vev{ \al_e\al_c\al_d } \nonumber\\
&&+ \Vev{ \al_a\al_b\al^e } 
   \Vev{ \al_e\al_c\al_d \int\!\phi_i^{(2)}\int\!\phi_j^{(2)} } \\
&&+ \Vev{ \al_a\al_b\al^e \int\!\phi_i^{(2)} } 
   \Vev{ \al_e\al_c\al_d \int\!\phi_j^{(2)} } 
   \mp\mbox{cyclic perms.} \nonumber
\end{eqnarray}

The first two terms on the right-hand side together equal half the
left-hand side. The derivation is similar to \eqref{randtermen}, with 
$\phi_i$ replaced by $\phi_i\int\!\phi_j^{(2)}$.\footnote{In general, 
we have the operator identity 
$Q\Bigl(\phi_i \e^{\,t^k\!\!\int\!\phi_k^{(2)}}\Bigr) = 
t^j[\phi_j,\phi_i] \e^{\,t^k\!\!\int\!\phi_k^{(2)}}$. This reflects 
the well-known fact that the BRST operator is deformed by the bracket 
with $t^j\phi_j$ \cite{verl}. The extra integral should be performed 
on a (regularized) disc inside the contour used for the action 
of the BRST operator. Therefore we do not get deformed star products 
at the right-hand side of \eqref{randtermen}, but only the first two terms 
in \eqref{brack}.}
The bracket is therefore given by
\begin{equation}\label{bracket}
\Vev{ \alpha_d\, [\Phi_i,\Phi_j](\al_a,\al_b,\al_c) } = 
\Vev{ \al_a \al_b \al_e \int\!\phi_i^{(2)} } 
 \Vev{ \al^e \al_c \al_d \int\!\phi_j^{(2)} }
 \mp\mbox{cyclic~perms.}
\end{equation}
With extra boundary insertions, the derivation goes along the same lines, 
with both sides acquiring extra contributions.
This gives \eqref{haakje}. Finally we remark that the factorisation
can be generalised to the higher relations of the $A_\infty$ algebra,
giving generalised Maurer-Cartan equations.

\subsection{Deformation Complex:  Abstract Setting}

At this point it is useful to examine the more abstract relation
between deformations and Hochschild complex. The deformation algebra
$\Def^*(A)$ of any algebra is a graded Lie algebra, formed by the
possible deformations of the structure. These comprise not only the
actual deformations, but also symmetries and obstructions to
deformations. They always contain a trivial part equal to the algebra
itself, which represents the shift of the origin in the space of
operators. The rest consists of the various deformations of
multilinear maps, contained in the Hochschild complex, with the
grading shifted by 1. That is,
\begin{equation}
\Def^p(A) = A^{p} \oplus \Hoch(A)^{p+1}
 = A^{p} \oplus \bigoplus_{n\geq0} \Hom(A^{\otimes n},A)^{p+1-n}.
\end{equation}
The superscript is the grading. The shift in grading by one is 
actually a convention, as the degree in the mathematical deformation 
complex is always such that the actual (physical) deformations have degree 
one, and the gauge symmetries degree zero. The physical ghost number is 
more in line with the degree in the Hochschild complex. For example, the 
physical deformations are contained in the Hochschild complex at degree two, 
which is exactly the ghost number of the physical bulk operators (in the 
zeroth descendants form). The extra shift of $n$ in the 
last identification is related to the fact that exactly that number 
of boundary operators in the corresponding correlators $\Phi_{ia_0\ldots a_n}$ 
are descendants. 
\begin{table}[t]
\begin{center}
\begin{tabular}{|c|c|c|}
\hline
$\Def^0(A)$ & Symmetry & CS \\
\hline
$A^0$       & $\delta\alpha = \{Q,\lambda_0\}$   & $\delta A = d_A\lambda_0$ \\
$A^1$       & $\delta\phi   = \{Q,\lambda_1\}$   & $\delta B = d_A\lambda_1$ \\
$\End(A)^0$ & $\delta\alpha = \lambda(\alpha)$ & $\delta A = \lambda(A)$ \\
\hline \hline
$\Def^1(A)$ & Deformation & CS \\
\hline
$A^1$                     & $\delta\alpha = \alpha_1$        & $\delta A = \lambda_1$ \\
$A^2$                     & $\delta F = \varphi_2$              & $\delta F = b_2$ \\
$\End(A)^1$               & $\delta Q = \oint\!\varphi_2^{(1)}$ & $\delta d_A f= a_1\wedge f$ \\
$\Hom(A^{\otimes 2},A)^0$ & $\delta * = \int\!\varphi_2^{(2)}$  & $f\mathbin{\delta*}g = \theta(f,g)$ \\ 
\hline \hline
$\Def^2(A)$ & Obstruction & CS \\
\hline
$A^2$ & $\{Q,\alpha\}=0$ & E.o.m. \\
$A^3$ & $\{Q,F\}=0$ & Bianchi \\
$\End(A)^2$ & $Q^2=0$ & Derivation \\
$\Hom(A^{\otimes 2},A)^1$ & $[Q,*]=0$ & Leibniz \\
$\Hom(A^{\otimes 3},A)^0$ & $*^2=0$ & Associativity \\
\hline 
\end{tabular}
\end{center}
\caption{Components of $\Def^*(A)$ and their interpretations as gauge symmetries, 
deformations and obstructions. $\delta$ denotes a variation. The structure 
of the open string algebra is determined by the operations $F$, $Q$, and $*$.}
\label{tb:def}
\end{table}

The elements of degree 0 correspond to the
(gauge) symmetries of the algebra. The actual physical deformations
are contained in $\Def^1$. The degree 2 part, $\Def^2$, consists of
the obstructions to deformations. They express the conditions that
should be satisfied. Physically, they correspond to classical
equations of motion of the background. 
Some important components of $\Def^*(A)$ are summarised in Table
\ref{tb:def}. The last column shows the interpretation in the
Chern-Simons gauge theory, which can be viewed as an open string theory. 
This will be discussed in some more detail in Section \ref{abmodel}. 
The BRST operator is the covariant derivative. 
In general, after a deformation, we also have to
include an element $F$ in the algebra, which can be interpreted as a degree 
zero component $d_0$ of the total differential on the $A_\infty$ algebra 
(compare Appendix \ref{hom}). On-shell this element is clearly zero. 
In CS, it has the interpretation of the field strength.
The deformation in $A^2\subset\Def^1(A)$ by the degree two element
corresponds to a deformation in the action by a shift of this element
$F$.  In the gauge theory it is similar -- and actually directly
related in a topological limit -- to the addition of the two-form
field $B$ to the total field strength. Its symmetry is contained in
$A^1\subset\Def^0(A)$.

In the cohomology, the deformation $\delta\alpha = \alpha_1$ is cancelled 
by the (gauge)symmetry $\delta\phi_2=\{Q,\lambda_1\}$, taking 
$\lambda_1=\alpha_1$. The differential of the deformation complex maps 
$A^1\subset\Def^0(A)$ identically onto $A^1\subset\Def^1(A)$, 
so that in cohomology they cancel each other.

\subsection{A Special Case}

As a special case, consider $A=C^\infty(M)$, the case of Kontsevich
and of Seiberg and Witten (functions or polynomials). The Hochschild
complex shifted by 1 is  the complex of polydifferential operators
on $M$; the star product corresponds to a bidifferential $A^{\otimes 2} \rightarrow A$.\\ 
The Hochschild cohomology consists of the polyvector fields on $M$,
which have an independent differential Gerstenhaber structure, given
by a vanishing differential, wedge multiplication of polyvector
fields, and the Schouten-Nijenhuis bracket, which is the usual Lie
bracket on vector fields, and is extended to polyvector fields by the
Leibniz rule, $ [\theta_1, \theta_2\wedge\theta_3] =
[\theta_1,\theta_2] \wedge\theta_3 + (-1)^{|\theta_1|(|\theta_2|-1)}
\theta_2\wedge [\theta_1,\theta_3]$.

A bivector field $\theta$ satisfying the Maurer-Cartan equation on
cohomology, $[\theta, \theta]=0$, corresponds to a Poisson structure
$\{ f,g \} = \theta^{ij} \del_i f \del_j g$ on $M$ and, when
invertible, to a 2-form $B=\theta^{-1}$ satisfying $dB=0$.\footnote{In
components of $\theta$, $\theta^{[il}\del_l \theta^{jk]} =0$.} Indeed, 
$B$ is the physical $B$-field, a closed string state.

In this special case, there is a correspondence between deformers (in
the Hochschild cohomology) and deformations (in the Hochschild
complex) preserving the algebra structure, so that the brackets, and
hence the Maurer-Cartan equations, are mapped onto each other. This
has been shown for the differential graded Lie structure by Kontsevich
\cite{kon1} and for the differential graded Gerstenhaber structure by
Tamarkin \cite{tam}. The correspondence, whose existence is captured
by the formality theorem, can be given by the perturbative expansion
of the Cattaneo-Felder model \cite{cafe}.

\section{Operad Formulation and Deligne Theorem}
\label{ope}

The algebraic structure of the Hochschild complex is encoded in the
statement that it is an algebra over the `operad of little discs' \cite{kon2}. 
This operad is closely related to two-dimensional CFTs.

\subsection{Little Discs}

Operads are mathematical devices to describe multilinear operations.
They are certain collections $O(n)$, with actions of the symmetry
group $S_n$ acting on them. They form a useful tool to study (local)
quantum field theories, as there collections of operators having a
particular behaviour under interchanges are associated to a point in
some topological space. This has been recognised in the literature
\cite{ksv,stasheff}.

The most relevant operad for us, and in general for a local perturbative 
quantum field theory, is the operad $C_d$ of little $d$-discs. Here $d$ 
is any nonnegative integer, which stands for the (spatial) dimension. 
It is a topological operad consisting of collections $C_d(n)$ of 
$n$ $d$-dimensional discs (holes) $D_i$, $i=1,\ldots,n$, inside the 
standard disc 
\begin{equation}
D_0 := \Bigl\{(x_1, \ldots , x_d) \in \mathbb{R}^d \Bigm| x_1^2 + \ldots + x_d^2 \leq 1\Bigr\}.
\end{equation}
There is a natural action of the symmetric group $S_n$ on $C_d(n)$, given by
the permutation of the indices of the little discs $D_i$.
The first two collections are given by $C_d(0) = \emptyset$ and 
$C_d(1) = \mbox{point} =\opname{id}_{C_d}$.

On these collections of discs, there are natural compositions 
\begin{equation}
C_d(k)\times C_d(n_1)\times\cdots C_d(n_k)\rightarrow C_d(n_1+\cdots +n_k),
\end{equation}
given by gluing $k$ discs inside a disk with $k$ holes. Being a bit 
more precise, let $G_d$ be the $(d+1)$-dimensional Lie group acting on $\R^d$ 
by affine transformations $u \mapsto \lambda u + v$ where $\lambda \in 
\R$ and $v \in \R^d$. The composition law is obtained 
by applying elements from $G_d$ associated with discs $D_i$, $i=1, 
\ldots, k$ in the configuration in $C_d(k)$ to configurations in all 
$C_d(n_i)$, $i=1, \ldots, k$, and putting the resulting configurations 
together. 

For $d=2$ we can naturally relate this to the composition of closed 
string diagrams. The little discs $D_i$ are the holes in which the 
(regularised) local operators are inserted, while the standard disc 
$D_0$ is to be thought of as the outgoing state at infinity. 
The affine group $G_2$ is the subgroup of the conformal group that 
remains after the choice of the standard disc. 

For technical reasons, one also introduces the operad 
$\func{Chains}(C_d(n))$ of chain complexes on the
topological operad $C_d$. Its cohomology is the homology operad
$H_*(C_d)$. For the precise definition, see \cite{kon2}.

\subsection{$d$-Algebras}

Next, to define a local QFT on a topological space, one has to
introduce an algebra of operators, which are to be inserted in the
holes.  The mathematical objects related to this are $d$-algebras,
which are related to the above operad. 

An {\it algebra over an operad} $O$ of vector spaces (complexes) is a
vector space (complex) $V$ provided with a morphism of operads
\begin{equation}
O(n) \rightarrow \Hom(V^n, V), \quad n \geq 1.
\end{equation}
We can also write this as a sequence of maps
\begin{equation}\label{opmap}
O(n) \times V^n \rightarrow V, \quad n \geq 1,
\end{equation}
satisfying certain compatibility conditions \cite{ksv}. One also says 
that \emph{the operad $O$ acts on $V$}. For the operad of 
little discs, \eqref{opmap} can be thought of as a collections 
of holes connected to operators in a linear space $V$, which map to 
an outgoing state in $V$ (at the fixed point at infinity). 

A $d$-algebra is an algebra over the operad $\func{Chains}(C_d)$ in
the category of complexes. Because $C_0$ is a point, a 0-algebra is a
complex. One can show that any 1-algebra carries a natural structure
of an $A_\infty$-algebra.  A $d$-algebra can thus be characterised by
a vector space $V$ together with maps in $\Hom(V^{\otimes n},V)$,
possessing a natural action of the groups $S_n$, and endowed with
composition laws compatible with the structure of the operad $C_d$.
$\func{Chains}(C_d(n))$ can be shown to be quasi-isomorphic to its
cohomology, $H_*(C_d)$, endowed with zero differential.\footnote{A
  quasi-isomorphism between objects of a given structure is a morphism
  with respect to this structure, inducing an isomorphism of
  cohomologies.} $d$-algebras are thus essentially algebras over
$H_*(C_d)$, and these can be classified. Up to homotopy, an algebra
over $H_*(C_d)$ is \cite{kon2}
\begin{itemize}
\item a complex if $d=0$,
\item a differential graded associative algebra if $d=1$,
\item a differential graded ``twisted'' Gerstenhaber algebra with 
bracket of degree $1-d$ for $d\geq 3$ even,
\item a differential graded ``twisted'' Poisson algebra with bracket 
of degree $1-d$ for $d\geq 3$ odd.
\end{itemize}
Physically, these characterisations are quite natural. For
$d=0$ we should think of instantons; the operators for an instanton
form a vector space. For a point particle, or $d=1$, we also know that
the operators in the quantum mechanics form an associative algebra.
The product can be thought of as applying two consecutive perturbations
at different times, and measuring the outcome at infinite time 
(having the topology of an interval with two holes). We mentioned
already that closed string theory ($d=2$) has the
structure of a Gerstenhaber algebra.

\subsection{A Realisation of the Deligne Theorem}

The Deligne theorem, proved by Kontsevich and Soibelman
\cite{kon3}, states that there exists a natural action of the operad
$\func{Chains}(C_2)$ on the Hochschild complex $C^*(A,A)$ of any
associative or $A_\infty$-algebra $A$, i.e. that $\Hoch(A) = C^*(A,A)$
is a 2-algebra.

This can be interpreted precisely as our statement that the deformations of 
the open string algebra form a consistent (tree level) closed string theory. 
The Deligne theorem means that there are maps $\Hom(C^*(A,A)^{\otimes n},C^*(A,A))$ 
whose compositions are compatible with the ones in the operad of little discs. 
This is the structure of the open strings we found: the
structure of $C_2$ is formed by the discs with holes, and these have
composition laws given by gluing discs in holes. The maps in
$\Hom(V^{\otimes n},V)$ can be seen as a map from a state at the holes
of a disc with $n$ holes to a state at the boundary of the disc.  Note
that one can view the boundary of the disc as a state at infinity.
This is the structure of closed string operators. The Deligne
theorem can then be interpreted as saying that the element of
$\Hoch(A)$ can be inserted at a boundary. These are precisely
the states $|\alpha_{a_0}\alpha_{a_1}\cdots\alpha_{a_n}\rangle$, which
we should put at the boundary of a hole.

\section{Gauge Theory Realisations}
\label{abmodel}

In this sections we study two gauge theory realisations of the above
picture, Chern-Simons and holomorphic Chern-Simons theory. They can be
viewed as open string theories corresponding to the topological
A-model and B-model respectively \cite{witcs}. The reader should be
warned that some of the results of this section are somewhat
preliminary, and only a sketch of the structure is presented.
More details will be presented in a forthcoming paper \cite{homaHCS}.

\subsection{Chern-Simons Theory}

Chern-Simons was obtained as the effective field theory for a
topological open string theory by Witten \cite{witcs}.  
It is the open string sector of the topological A-model, which
describes the K\"ahler moduli space of a Calabi-Yau manifold. The
Chern-Simons theory lives on a (special) Lagrangian 3-cycle, or
supersymmetric D3-brane\footnote{We work in Euclidean 
space, and count only dimensions lying inside the Calabi-Yau.}, 
embedded in the Calabi-Yau manifold. We know that the theory living on a 
D-brane is a gauge theory. On this three-cycle lies a flat gauge bundle 
$E$, a critical point of CS.

The closed string A-model has operators corresponding to the
deformations of the K\"ahler form and of the $B$-field on the
Calabi-Yau. The BRST cohomology can be identified with the Hodge
cohomology $H^{*,*}(M)=H^*(M,\C)$ of the Calabi-Yau $M$. The
total BRST operator can be identified with the $d$-operator.

The tangent space of the open string model, spanned by the boundary
operators $\alpha_a$, consists of the deformations of the connection
$A$ of the gauge bundle $E$ on the 3-cycle, and of the transverse scalars
in the adjoint. In the rest we will ignore these scalars, and
concentrate on the pure gauge theory. The BRST operator of the open
string becomes the covariantised form of the closed string BRST
operator, $Q = d_A$. Similarly, the operator $G$ is given by
$G=d_A^*$. The derivation condition $Q^2=0$ translates to $F=d_A^2=0$,
expressing the fact that the gauge bundle should be flat. The tangent
space is at any point isomorphic to $\Omega^{1}(M,\End(E))$, the space
of adjoint valued $1$-forms.  We will denote the variations by
$\delta A$.  The corresponding zero-form vertex operator and its
descendant are given by $\alpha = \delta A_{\mu}(X)\chi^{\mu}$ and
$\alpha^{(1)} = \delta A_{\mu}(X)\dot X^{\mu}$ respectively, where 
$\chi^\mu$ denote the fermions that are twisted to worldsheet scalars, 
and the dot denotes the derivative with respect to a coordinate along the
boundary.  The BRST cohomology of the open string A-model is therefore
given by $H^*(M,\End(E))$.

Using this correspondence of operators and (variation of the) gauge field, 
we can  express all  correlators of CS in terms of an effective field theory, 
living on the worldvolume of the D-brane. 
The effective action is given by \cite{witcs}
\begin{equation}
S = \int_M \Tr\biggl(\frac{1}{2}A\wedge d A + \frac{1}{3}A\wedge A\wedge A\biggr) 
+ \mbox{instanton~corrections}.
\end{equation}
The corrections are due to string instantons. Due to these, the theory is 
not precisely CS, but a deformed one. The deformation parameter is basically the 
complexified K\"ahler form. 
For example, the structure constants of the open string algebra are 
given in terms of this effective field theory by 
\begin{equation}
F_{abc} = \int_M \Tr\Bigl(\delta_a A\wedge\delta_b A\wedge\delta_c A \Bigr)+\cdots 
\end{equation}
(the dots denote the instanton corrections). The first mixed correlator, 
for a bulk operator $\phi_i$ corresponding to a $(1,1)$-form $\theta_i$, 
is given by 
\begin{eqnarray}
\Phi_{ia} = \Vev{ \phi_i\alpha_a } &=& 
 \Vev{ (\theta_i\chi\tilde\chi)(\chi\delta_a A(X)) } + \cdots \nonumber\\
 &=& \int_M \theta_i\wedge\Tr\Bigl(\delta_a A\Bigr) + \cdots.
\end{eqnarray}
Here we used the fact that $\chi$ has three zero modes, corresponding to 
the three longitudinal directions. 
For the higher mixed correlators, we have to be careful to insert the first descendants 
$\dot X\delta A$ for the extra boundary insertions. The $\dot X$ gives extra contributions 
due to the nonvanishing $X-\dot X$ OPE, which are easily calculated. We find 
for the lower mixed correlators
\begin{eqnarray}
\Phi_{iab} &=& \int_M \theta_i\wedge \Tr d_A^*\Bigl(\delta_a A\wedge\delta_b A \Bigr)+\cdots.\\
\Phi_{iabc} &=& \int_M \theta_i^{\mu\nu}\Tr\Bigl(\del_\mu\delta_a A\wedge\del_\nu\delta_b A\wedge\delta_c A \Bigr)+\cdots.
\end{eqnarray}
The last expression can be seen as the noncommutative variation 
of the product. 

The Hochschild cohomology of the open string algebra $A=H^*(M,\End(E))$ is given by 
\begin{equation}
H^*(\Hoch(A)) = H^*(M).
\end{equation}
The important part is the physical deformation, $H^1(\Def(A))=H^2(\Hoch(A)) = H^2(M)$. 
The realisation is by the noncommutative deformation of the gauge theory. 
This seems to be only half of the closed string moduli space, corresponding 
to the $B$-field. Comparing to the closed string, for which the ground ring is given by 
$H^*(M,\C)$, the contribution of the K\"ahler form deformation is missing. 
It may appear if we study the full supersymmetric theory, including also the sector 
corresponding to the transverse scalars. Also note that 
the K\"ahler form only appears in the instanton corrections, as the pure CS part 
does not depend on the metric. On the other hand, we considered only the perturbative 
part of the story here. 

The deformation complex is given in more detail in the last column of Table \ref{tb:def}.  First, the symmetries
are in $\Def^0(A) = A^0\oplus A^1\oplus\End(A)^0$. The first term is the usual gauge
symmetry of the gauge field by a zero-form. The second term is the
gauge symmetry of the $B$-field, which can be added in the action to
the field strength. The third term is the invariance under
reparametrisations of the gauge field.

The physical deformations are given by 
$\Def^1(A) = A^1\oplus A^2\oplus\End(A)^1\oplus\Hom(A^{\otimes 2},A)^0$. 
The first term is the shift of the gauge field. This deformation should 
be identified with the gauge transformation of the $B$-field. The second 
term represents the shift of the field strength by the $B$-field, which 
is a deformation of the action. The next term is the deformation of 
the covariant derivative by (the variation of) the gauge field, which is a 
map of the algebra (of forms) increasing the degree by one. The last 
term is the noncommutative deformation of the (wedge) product. 

Last, we discuss the obstructions. They are in $\Def^2(A) = A^2\oplus
A^3\oplus\End(A)^2\oplus\Hom(A^{\otimes 2},A)^1\oplus\Hom(A^{\otimes 3},A)^0$.  
The first term corresponds to the condition $d_A\delta
A=0$, which is the equation of motion for the variation of the gauge
field. The second one is the Bianchi identity, $d_AF=0$. The third is
the derivation condition on the covariant derivative, $d_A^2=0$.
Next, we have the Leibniz rule $d_A(fg) = (d_A)g+fd_Ag$.  The last
term then is the associativity of the product, $(f*g)*h = f*(g*h)$.
Apart from the first one, they correspond to the constraints in the
$A_\infty$ algebra.

\subsection{Holomorphic Chern-Simons Theory}

Holomorphic Chern-Simons theory, also introduced by Witten
\cite{witcs}, is the boundary theory related to the closed string
B-model, which is described by Kodaira-Spencer theory (KS) -- the
deformation theory of complex structures on a Calabi-Yau 3-fold $M$.
Much is known about KS (see e.g.  \cite{kosp,witmir,bece}), and also
HCS has been studied in some detail, though certainly not as much (see
e.g.\ \cite{gova,vafa,hopa}).

The operators in KS theory are built from the twisted fermions $\chi_\mu$ 
and $\eta^{\bar\nu}$. They are sections of the holomorphic tangent 
bundle $T_M$ and the anti-holomorphic cotangent bundle $\overline T_M^*$ 
to $M$, hence any element $\theta\in\Gamma(M,\ext[p]\overline T_M^*\otimes\ext[q]T_M)$ 
corresponds to an operator\footnote{Underlined indices denote multi-indices.} 
$\phi = \theta^{\vec{\mu}_p}_{\vec{\bar\nu}_q}(X)\chi_{\vec{\mu}_p}\eta^{\vec{\bar\nu}_q}$.
The BRST operator corresponds to the $Q=\bar\del$ operator on this space, and 
$G=\bar\del^*$. The BRST cohomology of KS is given by the spaces 
$H^{-p,q}(M) \equiv H^q_{\bar\del}(M,\ext[p]T_M) \simeq H^{3-p,q}(M)$.
The most important generators, corresponding to the actual moduli 
of KS, correspond to the forms contained in $H^{-1,1}(M)$. 

The open string sector of the B-model is given by 
holomorphic Chern-Simons theory (HCS) \cite{witcs}. The KS theory 
couples naturally to holomorphic bundles, or more generally, to
even-dimensional D-branes wrapped around holomorphic cycles in 
the Calabi-Yau with a holomorphic bundle on it. These are the natural 
boundary conditions which are invariant under the BRST operator 
of KS. They naturally describe D6-branes wrapped around the Calabi-Yau space. 
The BRST operator of the open string becomes the covariantised 
form of the closed string BRST operator, $Q = \bdA$, and the operator $G$ 
is given by $G=\bdA^*$. The derivation condition $Q^2=0$ 
translates to $F^{0,2}=\bdA^2=0$, expressing the fact that the 
bundle has to be holomorphic. The tangent space is at any point isomorphic 
to $\Omega^{0,1}(M,\End(E))$, the space of $(0,1)$-forms with values 
in the adjoint bundle. The corresponding zero-form vertex operator and its 
descendant are given by $\alpha = \delta A_{\bar\mu}(X)\eta^{\bar\mu}$ and 
$\alpha^{(1)} = \delta A_{\bar\mu}(X)\dot X^{\bar\mu}$ respectively. 
The BRST cohomology of the open string B-model is therefore given by 
$H^*_{\bdA}(M,\End(E))$.

The effective action is given by \cite{witcs}
\begin{equation}
S = \int_M \Omega \wedge\Tr
\biggl(\frac{1}{2}A\wedge\bar\del A + \frac{1}{3}A\wedge A\wedge A\biggr).
\end{equation}
The structure constants of the open string algebra are 
given in terms of this effective field theory by 
\begin{equation}
F_{abc} = \int_M \Omega \wedge\Tr
\Bigl(\delta_a A\wedge\delta_b A\wedge\delta_c A \Bigr).
\end{equation}
Some mixed correlators, with a $(2,1)$-form $\mu_i$ representing the 
variation of the three form $\Omega$ by a change of complex structure, 
are given by 
\begin{eqnarray}\label{pia}
\Phi_{ia} &=& \int_M \mu_i \wedge\Tr\Bigl(\delta_a A\wedge F\Bigr), \\
\label{piab}
\Phi_{iab} &=& \int_M \mu_i \wedge \Tr\Bigl(
  \delta_a A\wedge\dA\delta_bA\Bigr), \\
\label{piabc}
\Phi_{iabc} &=& \int\mu_i\wedge\Tr\biggl(
  \dA\delta_a A\wedge\frac{\bdA^*}{\bar\Delta_A}(\delta_b A\wedge\delta_c A)
  \biggr),
\end{eqnarray}
where $\bDA=\bdA\bdA^*+\bdA^*\bdA$ is the closed string Hamiltonian. 
As a consistency check, notice that as $\delta_b F^{1,1}=\dA\delta_b A$ we can view 
\eqref{piab} as a variation of \eqref{pia} -- in this case there is no 
ordering problem for the boundary operators. 
The presence in \eqref{piabc} of the single open string propagator, 
$\frac{b_0}{L_0} = \frac{\bdA^*}{\bDA}$, is related to the fact that it 
is can be calculated using a tree-level calculation of a four-point function, 
for which we can write down three Feynman diagrams, with a single internal line. 
Similar formulas can be written down for the higher mixed correlators. 

The general mixed correlators form the Hochschild cohomology 
of HCS. The open string algebra was given by $A=H^*_{\bdA}(M,\End(E))$, 
generated by the holomorphic coordinates $z^\mu$, and the fermionic zero-modes 
$\eta^{\bar\mu}$. For the closed string algebra, we had $H^{-p,q}(M)$, generated 
by $z^\mu$, $\eta^{\bar\mu}$, and $\chi_\mu$. 
From the mixed correlators we read off the correspondence 
$H^{q-1}(\Hom(A^{\otimes p+1},A))=H^q_{\bar\del}(M,\ext[p]T_M)$. 
This agrees with direct calculation: using the Hochschild-Kostant-Rosenberg theorem, 
we find that the Hochschild cohomology of $A$ should be generated by 
the generators of $A$, supplemented by generators $\chi_\mu$  
of (ghost) degree 1, and $\lambda_{\bar\mu}$ of degree zero.\footnote{In 
terms of $\Hom(A^{\otimes n},A)$, one can think of them as the cohomology 
classes of the maps $\del/\del z^\mu$ and $\del/\del\eta^{\bar\mu}$, respectively.} 
In the cohomology, the two degree zero generators do not play much of a role.  
The two fermionic generators on the other hand precisely generate the bundles 
$T_M$ and $\overline T_M^*$, respectively.

\section{Conclusions and Outlook}
\label{concl}

In this paper we have expressed deformations of topological open
strings in terms of mixed correlators, identified them with the
Hochschild cohomology of the open string algebra, and shown that 
this Hochschild cohomology is aclosed string algebra. The mixed 
correlators \eqref{mixed} can beregarded as inner products 
$\Phi_{ia_0a_1\ldots a_n} = \vev{i|a_0a_1\ldots a_n}$, 
intertwining between a boundary state 
$\ket{a_0a_1\ldots a_n} \equiv\ket{\vec{a}}$ and
a bulk state $\ket{i}$. This gives an identification
\begin{equation}\label{dict}
\ket{i} \simeq \sum_{\vec a}\Phi_{i}{}^{\vec{a}}\ket{\vec{a}}.
\end{equation}
In the open string channel, open string states can mediate between boundary 
conditions, such as given by Chan-Paton indices. Denoting the vector space 
spanned by them by $B$, the open string state $\ket{\alpha_{\vec a}}$ 
is a single trace in $B$, so the above relation can be written in terms
of operators as 
\begin{equation}\label{trace}
\phi_i\simeq \sum_{\vec a}\Phi_{i}{}^{a_0a_1\ldots a_n}
\Tr\Bigl(\alpha_{a_0}\alpha_{a_1}^{(1)}\cdots\alpha_{a_n}^{(1)}\Bigr).
\end{equation}
We have identified the algebraic operations in the closed string
theory. The algebraic structure is that of a Gerstenhaber algebra, or
in the context of operads of little discs, that of a 2-algebra. This is a
physical realisation of the Deligne theorem, and it generalises the
Cattaneo-Felder model (function algebras). We have applied our
framework to the A- and the B-model.

It would be interesting to investigate if Kontsevich's formality
theorem holds for open string algebras, as it does for function
algebras.  The formality theorem states the existence of a
quasi-isomorphism between Hochschild complex and Hochschild
cohomology. In the case of function algebras, the essential argument
is that the Hochschild cohomology of the Hochschild cohomology is
trivial \cite{kon2}, i.e. that the algebra of deformations -- or
equivalently the closed string -- has only trivial deformations
itself.\footnote{This seems to contradict the WDVV construction of a
deformation of the closed string. However, that is a deformation of
the OPE algebra, while the relevant structure for the formality of
the Hochschild complex is the bracket, which is not deformed by
WDVV.} In our case it should first be shown that for every
infinitesimal deformation of the topological open string there is a
closed string state giving rise to this deformation. One expects a
full correspondence when the topological open/closed theory is
unitary. Subsequently, it should be shown that there are no affine 
invariant obstructions to the deformation of the deformation algebra.
In superstring theory, formality will probably not hold due to the 
affine invariant obstruction provided by the constant dilaton.

Inclusion of boundary conditions has no effect on the Hochschild
cohomology. This is due to the mathematical fact that the Hochschild
cohomology is invariant under Morita equivalence. That the closed
string theory does not depend on the choice of boundary conditions is
in agreement with this.

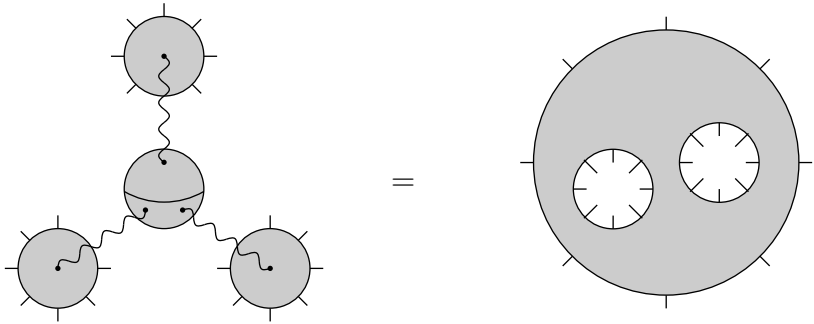
\begin{figure}
\begin{center}
\begin{picture}(290,120)
\put(0,0){
  \begin{picture}(120,120)(0,20)
    \Line(20,25)(20,20)
    \Line(5,40)(0,40)
    \Line(35,40)(40,40)
    \Line(30,30)(34,26)
    \Line(10,30)(6,26)
    \Line(10,50)(6,54)
    \Line(20,55)(20,60)
    \GOval(20,40)(15,15)(0){.8}
    \Line(100,25)(100,20)
    \Line(85,40)(80,40)
    \Line(115,40)(120,40)
    \Line(110,30)(114,26)
    \Line(90,30)(86,26)
    \Line(110,50)(114,54)
    \Line(100,55)(100,60)
    \GOval(100,40)(15,15)(0){.8}
    \Line(60,135)(60,140)
    \Line(45,120)(40,120)
    \Line(75,120)(80,120)
    \Line(50,130)(46,134)
    \Line(70,130)(74,134)
    \Line(70,110)(74,106)
    \Line(50,110)(46,106)
    \GOval(60,120)(15,15)(0){.8}
    \GOval(60,70)(15,15)(0){.8}
    \CArc(60,95)(30,240,-60)
    \Photon(20,40)(53,62)24
    \Vertex(53,62)1    
    \Vertex(20,40)1    
    \Photon(100,40)(67,62)24
    \Vertex(67,62)1    
    \Vertex(100,40)1    
    \Photon(60,120)(60,80)24
    \Vertex(60,80)1    
    \Vertex(60,120)1    
  \end{picture}
}
\put(150,50){=}
\put(190,0){
  \begin{picture}(100,100)
    \Line(60,110)(60,115)
    \Line(60,10)(60,5)
    \Line(10,60)(5,60)
    \Line(110,60)(115,60)
    \Line(95,95)(99,99)
    \Line(25,95)(21,99)
    \Line(95,25)(99,21)
    \Line(25,25)(21,21)
    \GOval(60,60)(50,50)(0){.8}
    \GOval(40,50)(15,15)(0)1
    \Line(40,65)(40,60)
    \Line(40,35)(40,40)
    \Line(25,50)(30,50)
    \Line(55,50)(50,50)
    \Line(51,61)(46,56)
    \Line(29,61)(34,56)
    \Line(29,39)(34,44)
    \Line(51,39)(46,44)
    \GOval(80,60)(15,15)(0)1
    \Line(80,75)(80,70)
    \Line(80,45)(80,50)
    \Line(65,60)(70,60)
    \Line(95,60)(90,60)
    \Line(91,71)(86,66)
    \Line(69,71)(74,66)
    \Line(69,49)(74,54)
    \Line(91,49)(86,54)
  \end{picture}
}
\end{picture}
\end{center}
\caption{An open string loop diagram.}
\label{osld}
\end{figure}

A further step would be to consider open string loop diagrams, which
can also be interpreted as disc diagrams connected by closed strings
(see Figure \ref{osld}), or comparing with \eqref{trace}, as
multi-trace operators. We could view the boundaries with open string
insertions as effective closed string operators, using the dictionary
\eqref{dict}. These generalised boundary operators however will
generally not be on-shell.  Hence an extension of the deformation
theory including off-shell closed string states is necessary. The
corresponding objects are homotopy Gerstenhaber or
homotopy BV algebras. Furthermore, one can ask the question whether
all external closed string insertions can be replaced by these
generalised boundary states, or under which conditions this can be
done. If this is the case, one can calculate all (tree-level) closed
string diagrams in terms of (multi-loop) open string diagrams.  This
is closely related to the open/closed string correspondence. 

An obvious generalisation is to higher $d$-algebras. The structure of
a $d$-algebra is rather similar to that of a perturbative
$d$-dimensional field theory. For higher $d$, the products are given
by the OPE and its higher homotopies, and the brackets are defined by
integrating descendants. The generalised Deligne theorem
\cite{kon2,kon3} would be a bulk-boundary correspondence; it states
the existence of a natural action (in a specific sense) of a
$(d+1)$-algebra on a $d$-algebra. For example, the relation between
Floer and Donaldson-Witten theories (see e.g.\ \cite{witdw}) can be
seen as an action of a $1$-algebra (quantum mechanics) on a
$0$-algebra (the configuration space of instantons). The action of a
3-algebra on a 2-algebra can be studied in a system of open M2-branes
ending on M5-branes, where deformations of closed strings might be
found.  In this case, deformations corresponding to constant 3-form
fields seem to correspond to a noncommutative version of loop space
\cite{bebe,bebe2,kasa,om}.  Work on the algebraic structure of such 
a system is in progress \cite{homa}.

Studying the generalisations may provide a new point of view on the
AdS/CFT-correspondence (AdS-gravity could be formulated in terms of a
$d$-algebra) and on RG-flow, especially in the context of \cite{kv}.
Another interesting relation is perhaps \cite{witten}.

\ack 

We like to thank Erik Verlinde, Robbert Dijkgraaf, Herman Verlinde,
Alberto Cattaneo, Jan de Boer and Anton van de Ven for useful
discussions and remarks. We are grateful to the organisers of the 1999
Workshop on Noncommutative Gauge Theory in Leiden for a very
stimulating week. We thank Princeton University for its hospitality.
C.H. is financially supported by the FOM foundation.

\appendix

\section{Integrability}
\label{wrd}

In this appendix, we use the Ward identity for the operator $G$ to
derive symmetry relations for the mixed correlators.  In the presence
of a boundary, the symmetry is $\SL(2,\R)$ and we have three real
vector fields $\xi=\xi(z)\del_z+\bar\xi(\bar z)\del_{\bar z}$.  The
Ward identity for $G$ can be written
\begin{eqnarray}
0 &=& \sum_m\xi(x_m)\Vev{ \prod_n\phi_{i_n}(z_n)
   \alpha_{a_1}(x_1)\cdots
   \alpha_{a_m}^{(1)}(x_m)\cdots\alpha_{a_r}(x_r)} \nonumber\\
 &&+ \sum_n\xi(z_n)\Vev{ \phi_{i_1}(z_1)\cdots\phi_{i_n}^{(1,0)}(z_n)
   \cdots\phi_{i_s}(z_s)\prod_m\alpha_{a_m}(x_m)} \\
 &&+ \sum_n \bar\xi(\bar z_n)\Vev{ \phi_{i_1}(z_1)\cdots\phi_{i_n}^{(0,1)}(\bar z_n)
   \cdots\phi_{i_s}(z_s)\prod_m\alpha_{a_m}(x_m)}.\nonumber
\end{eqnarray}
For multiple vector fields and forms, we use the convention that contractions are 
first applied between the vector field and the form that are closest to each other, 
e.g.\ $\xi_2\xi_1\alpha_a\alpha_b \equiv (\xi_1\cdot\alpha_a)(\xi_2\cdot\alpha_b)$. 
We write $\phi^{(2)}=\phi^{(1,1)}_{z\bar z}dz\wedge d\bar z$ and note that 
a contraction with a two-form is antisymmetric,
\begin{equation}
\bar\xi_2\xi_1\phi^{(2)} = \bar\xi_2\xi_1\phi^{(1,1)}_{z\bar z} 
 = -\xi_1\bar\xi_2\phi^{(1,1)}_{\bar zz}
 = -\xi_1\bar\xi_2\phi^{(2)}.
\end{equation}

In the case of one $\phi$ located in the bulk at $w$ and three
$\alpha$'s on the boundary at $x_1$, $x_2$ and $x_3$, we choose two
vector fields $\xi_1(z) = (z-x_2)(z-x_3)$ and $\xi_2(z) = (z-x_1)(z-x_3)$. 
Applying $\xi_1$,
\begin{equation}
0 = \xi_1(x_1)\Vev{ \phi\alpha^{(1)}\alpha^{(1)}\alpha }
 + \xi_1(w)\Vev{ \phi^{(1,0)}\alpha\alpha^{(1)}\alpha }
 + \bar\xi_1(\bar w)\Vev{ \phi^{(0,1)}\alpha\alpha^{(1)}\alpha },
\end{equation}
and then $\xi_2$, we find
\begin{eqnarray}
\xi_2(x_2)\xi_1(x_1)\Vev{ \phi\alpha^{(1)}\alpha^{(1)}\alpha }
 &=& -\xi_2(x_2)\xi_1(w)\Vev{ \phi^{(1,0)}\alpha\alpha^{(1)}\alpha }
     -\xi_2(x_2)\bar\xi_1(\bar w)\Vev{ \phi^{(0,1)}\alpha\alpha^{(1)}\alpha } \nonumber\\
 &=& \bar\xi_2(\bar w)\xi_1(w)\Vev{ \phi^{(2)}\alpha\alpha\alpha }
     -\xi_2(w)\bar\xi_1(\bar w)\Vev{ \phi^{(2)}\alpha\alpha\alpha }
\end{eqnarray}
\begin{equation}
\Vev{ \phi \al^{(1)} \al^{(1)}\al } = \biggl( \frac{\xi_1(w)}{\xi_1(x_1)}
\frac{\bar\xi_2(\bar w)}{\xi_2(x_2)} - \frac{\bar\xi_1(\bar w)}{\xi_1(x_1)}
\frac{\xi_2(w)}{\xi_2(x_2)} \biggr) \Vev{ \phi^{(2)} \al\al\al }.  
\end{equation}
Due to the conformal Ward identities, the correlators depend only 
on the anharmonic ratio $\zeta = \frac{(w-x_1)(x_2-x_3)}{(w-x_2)(x_1-x_3)}$, 
which satisfies 
\begin{equation}
\xi_i(x_i) \frac{\del\zeta}{\del x_i} + \xi_i(w) \frac{\del\zeta}{\del w}=0 \quad (i=1,2). 
\end{equation}
The result can be written as
\begin{equation}
\Vev{ \phi \al^{(1)} \al^{(1)}\al }
 = \biggl(\frac{\del\zeta}{\del w}\frac{\del\bar\zeta}{\del \bar w}\biggr)\inv
 \biggl(\frac{\del\zeta}{\del x_1}\frac{\del\bar\zeta}{\del x_2}
  -\frac{\del\zeta}{\del x_2}\frac{\del\bar\zeta}{\del x_1}\biggr)
   \Vev{ \phi^{(2)} \al\al\al }.
\end{equation}
Recognising a quotient of Jacobians, we conclude
\begin{equation}
\dint{dx_1 dx_2} \Vev{ \phi  \al^{(1)} \al^{(1)}\al } 
 = \dint{dwd\bar w} \Vev{ \phi^{(2)} \al\al\al }.
\end{equation}

Next we consider two $\phi$'s at $w$ and $v$, and one
$\alpha$ at $x$. We use two vector fields, $\xi_1(z) = z-x $ and 
$\xi_2(z) = (z-x)^2$.  Then we have ($i=1,2$)
\begin{equation}
0 = \xi_i(w) \Vev{ \phi^{(2)}\phi \al }  
  + \xi_i(v) \Vev{ \phi^{(0,1)}\phi^{(1,0)}\al } 
  + \bar\xi_i(\bar v) \Vev{ \phi^{(0,1)}\phi^{(0,1)}\al }
\end{equation}
the solution of which is
\begin{equation}
\pmatrix{ \vev{ \phi^{(0,1)} \phi^{(1,0)} \al } \cr
 \vev{ \phi^{(0,1)} \phi^{(0,1)} \al } }
  = \frac{1}{\xi_1(v)\bar\xi_2(\bar v) - \bar\xi_1(\bar v)\xi_2(v)} 
  \pmatrix{ \bar\xi_1(\bar v)\xi_2(w) - \xi_1(w) \bar\xi_2(\bar v)\cr 
  \xi_1(w)\xi_2(v) - \xi_1(v) \xi_2(w) }
  \Vev{ \phi^{(2)}\phi\al }.
\end{equation}
Interchanging $w$ and $v$ yields as the second equation,
\begin{equation}
\Vev{ \phi^{(0,1)} \phi^{(0,1)} \al } 
 = \frac{\xi_1(v)\xi_2(w) - \xi_1(w) \xi_2(v)}{\xi_1(w)\bar\xi_2(\bar w)
   - \bar\xi_1(\bar w)\xi_2(w)}\Vev{ \phi\phi^{(2)}\al },
\end{equation}
hence
\begin{equation}
\Vev{ \phi^{(2)}\phi\al }
 = - \frac{\xi_1(v)\bar\xi_2(\bar v) 
 - \bar\xi_1(\bar v)\xi_2(v)}{\xi_1(w)\bar\xi_2(\bar w) 
 - \bar\xi_1(\bar w)\xi_2(w)} \Vev{ \phi \phi^{(2)}\al }.
\end{equation}
Using the anharmonic ratio $\zeta = \frac{(v-x)(w-\bar w)}{(w-x)(v-\bar w)}$, 
satisfying similar relations as the one above, we can write
\begin{equation}
\Vev{ \phi^{(2)}\phi\al }
 = \biggl(\frac{\del\zeta}{\del v}\frac{\del\bar\zeta}{\del \bar v} 
  - \frac{\del\zeta}{\del\bar v}\frac{\del\bar\zeta}{\del v}\biggr)\inv
 \biggl(\frac{\del\zeta}{\del w}\frac{\del\bar\zeta}{\del \bar w}
  - \frac{\del\zeta}{\del \bar w}\frac{\del\bar\zeta}{\del w}\biggr)
  \Vev{ \phi \phi^{(2)}\al },
\end{equation}
so that
\begin{equation}
\dint{dwd\bar w} \Vev{ \phi^{(2)}\phi\al }
 = \dint{dv d\bar v} \Vev{ \phi \phi^{(2)}\al }.
\end{equation}

All these derivations remain true when there are extra insertions of 
$\int\!\phi^{(2)}$ and $\int\!\alpha^{(1)}$.

\section{Algebras up to homotopy}
\label{hom}

A \emph{homotopy associative} or \emph{$A_\infty$-algebra} can be defined in terms 
of a derivation $d$ acting on the tensor algebra $\CT A=\bigoplus_{n\geq0}A^{\otimes n}$ 
of a (graded) vector space $A$. The derivation is completely determined by 
the map from $\CT A$ to $A$. We denote the 
component of $d$ mapping the $n$th tensor product $A^{\otimes n}$ 
to $A$ by $d_n$. So we have $d=d_1+d_2+d_3+\cdots$. 
All $d_k$ are derivations in the sense that
\begin{equation}
d_k(a_1, \ldots, a_{k+n}) = \sum_{i=0}^{n} 
 (-1)^{i(n-1)}(a_1, \ldots, d_k(a_{i+1}, \ldots, a_{i+k}), \ldots, a_{k+n}).
\end{equation}
Furthermore, $d$ is a twisted differential, in the following sense. Considering 
the shifted algebra $sA=A[-1]$, called the suspension.\footnote{For 
an integer $k$, $[k]$ denotes a shift of the degree of a complex 
$C =\bigoplus_n C^n$ by $k$, that is $C[k]^n := C^{k+n}$; therefore, 
$sC^{n}=C^{n-1}$. Physically, the suspension corresponds to descent.}
The shifted maps $b_k=s\circ d_k\circ (s\inv)^{\otimes k}$ should form 
a differential on the suspension, i.e.\ $b^2=0$, of degree $-1$. 
This implies an infinite number of homogeneous relations for the $d_k$; 
for any $n\geq1$, 
\begin{equation}
\sum_{k+l=n+1}(-1)^{(k-1)l}d_k\circ d_l=0.
\end{equation}
The map $d_k$ has parity $k\mod 2\Z$. 
Explicitly, the first few relations read $d_1^2 = 0$, $d_1d_2 =
d_2d_1$, $d_2^2 = -d_1d_3-d_3d_1$, $d_2d_3-d_3d_2 =
-d_1d_4-d_4d_1$. These say that $d_1$ is a differential on $A$, $d_2$
is a product for which $d_1$ is a derivation, $d_3$ gives a correction
to the associativity of this product ($d_2^2$ is the associator),
etc. We often write $d_1(a)=\delta a$ and $d_2(a_1,a_2)=a_1 * a_2$.

\emph{Homotopy Lie} or \emph{$L_\infty$-algebras} are defined in a similar way.  
We also start with a (graded) space $A$. The only difference is
that everything should be (graded) anti-symmetric.  For example, the
tensor product of the algebra is replaced by the (graded) exterior
product, $\bigoplus_n\ext[n]A$, and the products $d_n$ are all
(graded) anti-commutative. They are called brackets, and $d_2^2=0$ is
the Jacobi identity for the Lie bracket.

A \emph{Gerstenhaber algebra} (G-algebra) is a $\mathbb{Z}$-graded
algebra with a graded commutative associative product $\cdot$ of degree 0 and
a bracket $[\cdot,\cdot]$ of degree $-1$ (the Gerstenhaber bracket), which is such 
that $A[-1]$ is a graded Lie algebra. The operations satisfy some relations. 
The obvious ones are the Leibniz rules for the product and the bracket, 
graded associativity for the product and graded Jacobi for the bracket. 
Furthermore, for any fixed element $\alpha$, the map $\beta\mapsto[\alpha,\beta]$ 
must be a graded derivation of the product.

A \emph{homotopy Gerstenhaber algebra} ($G_\infty$-algebra), is defined 
similarly to $A_\infty$ by first introducing a derivation/differential $d$ and 
then relaxing both associativity conditions. A $G_\infty$-algebra 
has contained in it both a commutative  $A_\infty$ and a $L_\infty$-algebra, 
according to the scheme 
\begin{equation}
\matrix{
G_\infty   &        & \rightarrow   & L_\infty   \cr
           & d      & [\cdot,\cdot] & [\cdot,\cdot,\cdot] & \cdots \cr
\downarrow & \cdot  \cr
\mathrm{comm.}~A_\infty & \{\cdot,\cdot,\cdot\} \cr
           & \vdots  
}.
\end{equation}
Horizontally, we have the Gerstenhaber bracket and the corresponding
higher bracket forming the $L_\infty$ structure, while vertically we
see the $A_\infty$ structure, based on the commutative product and the
higher products. They share the same $d$, so that on the
$d$-cohomology one always gets a Gerstenhaber algebra. An operad
definition of a homotopy Gerstenhaber algebra can be found in
\cite{kvz}.


\begin{thebibliography}{88}

\bibitem{codo} A.~Connes, M.R.~Douglas and A.~Schwarz,
\textit{Noncommutative Geometry and Matrix Theory: Compactification on Tori}, 
JHEP \textbf{9802} (1998) 003,
\texttt{hep-th/9711162}. 

\bibitem{cafe} A.S.~Cattaneo and G.~Felder, 
\textit{A path integral approach to the Kontsevich Quantisation Formula},
\texttt{math.QA/9902090}.

\bibitem{scho} V.~Schomerus
\textit{D-branes and Deformation Quantisation},
JHEP \textbf{9906} (1999) 030, 
\texttt{hep-th/9903205}.

\bibitem{seiwit} N.~Seiberg and E.~Witten, 
\textit{String Theory and Noncommutative Geometry}, 
JHEP \textbf{9909} (1999) 032,
\texttt{hep-th/9908142}.

\bibitem{kon1} M.~Kontsevich,
\textit{Deformation Quantisation of Poisson Manifolds, I},
\texttt{math.QA/9709180}.

\bibitem{kon2} M.~Kontsevich,
\textit{Operads and Motives in Deformation Quantisation},
\texttt{math.QA/9904055}.
\texttt{hep-th/9711162}.

\bibitem{zwie} B.~Zwiebach,
\textit{Closed String Field theory: Quantum Action and the BV Master Equation}, 
Nucl.\ Phys.\ \textbf{B390} (1993) 33-152,
\texttt{hep-th/9206084}.

\bibitem{kvz} T.~Kimura, A.A.~Voronov, G.J.~Zuckerman,
\textit{Homotopy Gerstenhaber Algebras and Topological Field Theory},
\texttt{q-alg/9602009}.

\bibitem{ksv} T.~Kimura, J.~Stasheff, A.A.~Voronov,
\textit{On Operad Structures of Moduli Spaces and String Theory},
\texttt{hep-th/9307114}.

\bibitem{stasheff} J.~Stasheff,
\textit{Closed string field theory, strong homotopy Lie algebras and 
the operad actions of moduli space},
\texttt{hep-th/9304061}

\bibitem{liz} B.H.~Lian and G.J.~Zuckerman,
\textit{Algebraic and Geometric Structures in String Backgrounds},
\texttt{hep-th/9506210}.

\bibitem{wizwi} E.~Witten and B.~Zwiebach,
\textit{Algebraic Structures and Differential Geometry in 2D String Theory},
Nucl.\ Phys.\ \textbf{B377} (1992) 55-112,
\texttt{hep-th/9201056}.

\bibitem{dvv} R.~Dijkgraaf, E.~Verlinde and H.~Verlinde,
\textit{Topological Open Strings in $d<1$},
Nucl.~Phys.~\textbf{B352} (1991) 59-86.

\bibitem{witeqn} E.~Witten,
\textit{Two-Dimensional Gravity and Intersection Theory on Moduli Space},
Surv.\ Diff.\ Geom.\ \textbf{1} (1991) 243-310.

\bibitem{zwieoc} B.~Zwiebach, 
\textit{Oriented Open-Closed String Theory Revisited},
Annals Phys.~\textbf{267} (1998) 193-248,
\texttt{hep-th/9705241}.

\bibitem{gerst} M.~Gerstenhaber and S.D.~Schack,
\textit{Algebras, Bialgebras, Quantum Groups and Alebraic Deformations},
Cont.\ Math.\ \textbf{134} (1992) 51.

\bibitem{stas} J.~Stasheff,
\textit{Homotopy Associativity of H-Spaces}, I and II,
Trans.\ Amer.\ Math.\ Soc.\ \textbf{108} (1963) 275 and 293.

\bibitem{wit} E.~Witten,
\textit{On Background Independent Open-String Field Theory},
\texttt{hep-th/9208027}.

\bibitem{kon4} M.~Kontsevich,
\textit{Feynman Diagrams and Low-Dimensional Topology}, 
First European Congress of Mathematics, 1990-1992 
Birkh\"auser (1993) 173.

\bibitem{verl} E.~Verlinde,
\textit{The Master Equation of 2d String Theory},
Nucl.~Phys.~\textbf{B381} (1992) 141-157,
\texttt{hep-th/9202021}.

\bibitem{tam} D.~Tamarkin, 
\textit{Another Proof of M.~Kontsevich Formality Theorem},
\texttt{math/9803025}.

\bibitem{witcs} E.~Witten,
\textit{Chern-Simons Gauge Theory as a String Theory},
\texttt{hep-th/9207094}.

\bibitem{kosp} K.~Kodaira and D.C.~Spencer, 
\textit{On Deformations of Complex Analytic Structures, I, II},
Annals of Math. II, \textbf{67} (1958) 45. 

\bibitem{witmir} E.~Witten,
\textit{Mirror Manifolds and Topological Field Theory},
in \textit{Mirror Symmetry I}, 121-160, 
S.T. Yau ed., International Press (1992)
\texttt{hep/th9112056}.

\bibitem{bece} M.~Bershadsky, S.~Cecotti, H.~Ooguri, and C.~Vafa,
\textit{Kodaira-Spencer Theory of Gravity and Exact Results 
for Quantum String Amplitudes},
Nucl.~Phys.~\textbf{B405} (1993) 279-304, 
\texttt{hep-th/9309140}.

\bibitem{gova} R.~Gopakumar and C.~ Vafa,
\textit{Topological Gravity as Large N Topological Gauge Theory},
Adv.~Theor.~Math.~Phys.~\textbf{2} (1998) 413-442,
\texttt{hep-th/9802016}.

\bibitem{vafa} C.~ Vafa,
\textit{Extending Mirror Conjecture to Calabi-Yau with Bundles}, 
\texttt{hep-th/9804131}.

\bibitem{hopa} C.M.~Hofman and J.-S.~Park, 
\textit{Sigma Models for Bundles on Calabi-Yau: A Proposal for 
Matrix String Compactifications},
Nucl.~Phys.~\textbf{B561} (1999) 125-156, 
\texttt{hep-th/9904150}. 

\bibitem{homaHCS} C.~Hofman and W.K.~Ma,
\textit{Kodaira-Spencer as a Deformation of Holomorphic Chern-Simons},
in preparation. 

\bibitem{kon3} M.~Kontsevich, Y.~Soibelman
\textit{Deformations of algebras over operads and Deligne's conjecture}
\texttt{math.QA/0001151}

\bibitem{witdw} E.~Witten,
\textit{Topological Quantum Filed Theory},
Comm.~Math.~Phys.~\textbf{117} (1988) 353.

\bibitem{bebe} E.~Bergshoeff, D.S.~Berman, J.P.~van der Schaar, and P.~Sundell
\textit{A Noncommutative M-Theory Five-brane},
\texttt{hep-th/0005026}.

\bibitem{bebe2} E.~Bergshoeff, D.S.~Berman, J.P.~van der Schaar, and P.~Sundell
\textit{Critical Fields on the M5-Brane and Noncommutative Open Strings},
\texttt{hep-th/0006112}.

\bibitem{kasa} S.~Kawamoto and N.~Sasakura,
\textit{Open Membranes in a Constant C-field Background and Noncommutative Boundary Strings}, 
\texttt{hep-th/0005123}.

\bibitem{om} R.~Gopakumar, S.~Minwalla, N.~Seiberg and A.~Strominger,
\textit{OM Theory in Diverse Dimensions},
\texttt{hep-th/0006062}.

\bibitem{homa} C.~Hofman and W.K.~Ma, 
\textit{Topological Model of M5-branes in a Background $C$-Field},  
in preparation.

\bibitem{kv} J.~Khoury and H.~Verlinde,
\textit{On Open/Closed String Duality},
\texttt{hep-th/0001056}.

\bibitem{witten} E.~Witten,
\textit{Duality Relations Among Topological Effects in String Theory},
\texttt{hep-th/9912086}.

\end{thebibliography}
\end{document}